\documentclass{aa}
\usepackage{txfonts}
\usepackage{graphicx}
\usepackage{natbib}
\usepackage{amssymb}
\usepackage{lscape}
\bibpunct{(}{)}{;}{a}{}{,}

\def\chandra{{\it Chandra~}}
\def\chandrak{{\it Chandra}}
\def\swift{{\it Swift~}}

\def\xmm{{\it XMM-Newton~}}
\def\xmmk{{\it XMM-Newton}}

\def\m31{{M~31}}
\def\msun{{$M_{\sun}$}}
\def\mwd{{$M_{\mbox{\tiny{WD}}}$}}

\def\mgc{HPH2009~}
\def\mgck{HPH2009}

\def\sqk{SQ2007}

\def\bol{{Bol~126~}}
\def\bolk{{Bol~126}}
\def\nova{{M31N~2010-10f~}}
\def\novak{{M31N~2010-10f}}
\def\gsss{{1E~1339.8+2837~}}

\newcommand{\nh}{\hbox{$N_{\rm H}$}~}
\newcommand{\hcm}[1]{$\times 10^{#1}$ cm$^{-2}$}

\newcommand{\ergs}[1]{$\times 10^{#1}$ \hbox{erg s$^{-1}$}}
\newcommand{\oergs}[1]{$10^{#1}$ erg s$^{-1}$}

\newcommand{\tpower}[1]{$\times 10^{#1}$}
\newcommand{\power}[1]{$10^{#1}$}

\newcommand{\ton}{$t_{\mbox{\small{on}}}$~}
\newcommand{\toff}{$t_{\mbox{\small{off}}}$~}
\newcommand{\eton}{t_{\mbox{\small{on}}}}
\newcommand{\etoff}{t_{\mbox{\small{off}}}}

\begin{document}

\title{Supersoft X-rays reveal a classical nova in the \m31 globular cluster \bolk\thanks{Partly 
   based on observations with \xmmk, an ESA Science Mission with instruments and contributions directly funded by ESA Member States and NASA}}

\author{M.~Henze\inst{1}
	\and W.~Pietsch\inst{1}
	\and F.~Haberl\inst{1}
	\and M.~Della Valle\inst{2,3}
	\and A.~Riffeser\inst{4}
	\and G.~Sala\inst{5,6}
	\and D.~Hatzidimitriou\inst{7,8}
	\and F.~Hofmann\inst{1}
	\and D.H.~Hartmann\inst{9}
	\and J.~Koppenhoefer\inst{1,4}
	\and S.~Seitz\inst{1,4}
	\and G.G.~Williams\inst{10}
	\and K.~Hornoch\inst{11}
	\and K.~Itagaki\inst{12}
	\and F.~Kabashima\inst{13}
	\and K.~Nishiyama\inst{13}
	\and G.~Xing\inst{14}
	\and C.H.~Lee\inst{15}
	\and E.~Magnier\inst{16}
	\and K.~Chambers\inst{16}
}

\institute{Max-Planck-Institut f\"ur extraterrestrische Physik, Postfach 1312, Giessenbachstr., 85741 Garching, Germany\\
	email: mhenze@mpe.mpg.de
	\and INAF-Napoli, Osservatorio Astronomico di Capodimonte, Salita Moiariello 16, I-80131 Napoli, Italy
	\and International Centre for Relativistic Astrophysics, Piazzale della Repubblica 2, I-65122 Pescara, Italy
	\and University Observatory Munich, Scheinerstrasse 1, 81679 M\"unchen, Germany
	\and Departament de F\'isica i Enginyeria Nuclear, EUETIB, Universitat Polit\`ecnica de Catalunya, c/ Comte d'Urgell 187, 08036 Barcelona, Spain
	\and Institut d'Estudis Espacials de Catalunya, c/Gran Capit\`a 2-4, Ed. Nexus-201, 08034, Barcelona, Spain
	\and Department of Astrophysics, Astronomy and Mechanics, Faculty of Physics, University of Athens, Panepistimiopolis, GR15784 Zografos, Athens, Greece
	\and Foundation for Research and Technology Hellas, IESL, Greece
	\and Department of Physics and Astronomy, Clemson University, Clemson, SC 29634-0978, USA
	\and Steward Observatory, 933 North Cherry Avenue, Tucson, AZ 85721, USA
	\and Astronomical Institute, Academy of Sciences, CZ-251 65 Ond\v{r}ejov, Czech Republic
	\and Itagaki Astronomical Observatory, Teppo-cho, Yamagata 990-2492, Japan
	\and Miyaki-Argenteus Observatory, Miyaki-cho, Saga-ken, Japan
	\and Xingming Observatory, Mt. Nanshan, Urumqi, Xinjiang, China
	\and Graduate Institute of Astronomy, National Central University, Jhongli 32001, Taiwan
	\and Institute for Astronomy, University of Hawaii at Manoa, Honolulu, HI 96822, USA
}

\date{Received ? / Accepted ?}

\abstract
{Classical novae (CNe) represent the main class of supersoft X-ray sources (SSSs) in the central region of our neighbouring galaxy \m31. Only three confirmed novae and three SSSs have been discovered in globular clusters (GCs) of any galaxy so far, of which one nova and two SSSs (including the nova) were found in \m31 GCs.}
{To study the SSS state of CNe we carried out a high-cadence X-ray monitoring of the \m31 central area with \xmm and \chandrak. This project is supplemented by regular optical monitoring programmes at various observatories.}
{We analysed X-ray and optical monitoring data of a new transient X-ray source in the \m31 GC \bolk, discovered serendipitously in \swift observations. Our optical data set was based on regular \m31 monitoring programmes from five different small telescopes and was reduced using a homogeneous method. Additionally, we made use of Pan-STARRS~1 data obtained during the PAndromeda survey. We extracted light curves of the source in the optical and X-rays, as well as X-ray spectra.}
{Our observations reveal that the X-ray source in \bol is the third SSS in an \m31 GC and can be confirmed as the second CN in the \m31 GC system. This nova is named \novak. Its properties in the X-ray (high black-body temperature, short SSS phase) and optical (relatively high maximum magnitude, fast decline) regimes agree with a massive white dwarf (\mwd~$\gtrsim 1.3$~\msun) in the binary system. Incorporating the data on previously found (suspected) novae in \m31 GCs we used our high-cadence X-ray monitoring observations to estimate a tentative nova rate in the \m31 GC system of 0.05 yr$^{-1}$ GC$^{-1}$. An optical estimate, based on the recent 10.5-year WeCAPP survey, gives a lower nova rate, which is compatible with the X-ray rate on the 95\% confidence level.}
{Although still based on small-number statistics, there is growing evidence that the nova rate in GCs is higher than expected from primordial binary formation and under conditions as in the field. Dynamical binary formation and/or additional accretion from the intracluster medium are possible scenarios for an increased nova rate, but observational confirmation for this enhancement has been absent, so far. Regular X-ray monitoring observations of \m31 provide a promising strategy to find these novae.}


\titlerunning{X-rays reveal second \m31 globular cluster nova}

\maketitle

\section{Introduction}
\label{sec:intro}
%
Classical novae (CNe), a subtype of cataclysmic variables (CVs) showing luminous optical outbursts \citep[see e.g.][]{2008clno.book.....B}, are rarely detected in globular clusters (GCs). Only three such discoveries are known: in the Galactic GC M\,80 \citep[nova T~Sco;][]{1860AN.....53..293L,2010ApJ...710..332D}, in a GC of the elliptical galaxy M\,87 \citep{2004ApJ...605L.117S}, and in the GC Bol~111 of our large neighbour galaxy \m31 \citep[][hereafter \sqk, \mgck]{2007ApJ...671L.121S,2009A&A...500..769H}. \citet{2004ApJ...605L.117S} argued that a fourth candidate, nova 1938 in the Galactic GC M\,14, was not a genuine GC nova.

An equally rare event is the discovery of a supersoft X-ray source \citep[SSS;][]{1991Natur.349..579T,1991A&A...246L..17G} in a GC; only three such objects have been found to date. The first was the transient SSS \gsss in the Galactic GC M\,3 (NGC 5272) \citep{1995A&A...300..732V,1999PASJ...51..519D}. This source was subsequently identified as a CV with unusual features \citep[][]{2004ApJ...611..413E,2011ApJ...732...46S} and might not fit into the classical picture  of SSSs as nuclear burning white dwarfs (WDs) \citep[see][and references therein]{1997ARA&A..35...69K}. The two other SSSs have been reported in \m31 GCs (\mgck).

Interestingly, one of these two \m31 SSSs was identified by \mgc with the GC nova discovered by \sqk. Classical novae have been found to constitute the majority of SSSs in the central region of \m31 \citep{2005A&A...442..879P}. Recently, \citet{2011A&A...533A..52H} published a catalogue of 60 novae in \m31 with a soft X-ray counterpart, a number significantly higher than for any other galaxy, including the Milky Way \citep[$< 30$; see][]{2011ApJS..197...31S}.

The SSS emission in CNe is believed to be a signature of stable hydrogen burning in accreted material on the surface of the WD that is not ejected during the nova outburst \citep{1974ApJS...28..247S,2005A&A...439.1061S}. Nova models describe that hydrogen-rich, degenerate matter accumulates on the WD surface until a thermonuclear runaway leads to a violent ejection of the hot envelope \citep[e.g.][]{1989clno.conf...39S}. This causes a strong rise in optical luminosity (on average 9-12 mag) within time scales of hours to days: the optical nova outburst. The underlying SSS becomes observable when the expansion of the ejected envelope reduces its opacity sufficiently \citep{2002AIPC..637..345K}. This time scale is defined here as the \textit{SSS turn-on time}, in agreement with earlier papers and theoretical work \citep[e.g.][]{2006ApJS..167...59H,2010ApJ...709..680H}, and should not be confused with the onset of the stable nuclear burning shortly after the outburst. As soon as its hydrogen fuel is exhausted, the SSS disappears. This time scale, the \textit{SSS turn-off time}, mainly depends on the amount of hydrogen left on the WD surface after the outburst \citep[][]{2005A&A...439.1061S}. For massive WDs, the expected SSS duration is very short \citep[$<$ 100~d;][]{1998ApJ...503..381T,2005A&A...439.1061S,2010ApJ...709..680H}.

It seems surprising that among the more than 900 nova candidates known in \m31 to date\footnote{August 2012: see the MPE online catalogue at\\ http://www.mpe.mpg.de/$\sim$m31novae/opt/m31/index.php} there is only a single GC nova. However, this might be explained by the fact that almost all optical surveys for CNe in \m31 that were conducted in the past \citep[see][and references therein]{2001ApJ...563..749S,2008A&A...477...67H} searched for suddenly appearing objects that were not visible before and fade back to invisibility in days to weeks. This condition is certainly not fulfilled by CNe in relatively bright GCs, where the optical background light of the GC itself makes a photometric discovery of a nova outburst much more complicated.

The first search specifically for novae in (54) \m31 GCs was carried out by \citet{1990PASP..102.1113C} based on H$\alpha$ data obtained by \citet{1987ApJ...318..520C,1990ApJ...356..472C}. Another pioneering work was the first, and so far only, spectroscopic survey by \citet{1992BAAS...24.1237T}, who monitored more than 200 \m31 GCs over an effective survey time of one year. Both studies did not detect any nova eruptions and reported upper limits on the nova rate in the \m31 GC system that were below the tentative rate later found by \mgc from their X-ray data.

This circumstance led \mgc to note that ``the detection of supersoft emission from a hydrogen-burning post-nova atmosphere is not affected [by the light of the GC]'' and to speculate that ``the connection of CNe to SSSs in X-rays provides a useful possibility to detect CNe in GCs''. In the present paper, we make use of this connection and describe the second nova found in an \m31 GC, which was first discovered as an SSS. The nova was detected in the GC \bol and given the name \novak. Section \ref{sec:obs} provides detailed information on our X-ray and optical data sets. Results are presented in Sect.\,\ref{sec:results} and discussed in Sect.\,\ref{sec:discuss} together with implications on the \m31 GC nova rate.

\section{Observations and data analysis}
\label{sec:obs}
%

\subsection{X-ray observations}
\label{sec:obs_xray}
The new X-ray source in the \m31 GC \bol was discovered serendipitously by \citet{2010ATel.3013....1P} during our \swift X-ray telescope \citep[XRT;][]{2005SSRv..120..165B} target of opportunity monitoring observations of the recurrent nova M31N~1963-09c \citep[see e.g.][and Henze et al. in preparation]{2010ATel.3001....1P,2010ATel.3038....1P}. Additional \swift observations followed the light curve of the object until the beginning of our regular X-ray monitoring \citep{2010AN....331..187P}\footnote{http://www.mpe.mpg.de/$\sim$m31novae/xray/index.php} of the \m31 central region with the telescopes \xmm and \chandra (PI: W. Pietsch). This programme used \xmm with the European Photon Imaging Camera \citep[EPIC;][]{2001A&A...365L..18S,2001A&A...365L..27T} as its primary instrument, while \chandra was operated with the High-Resolution Camera imaging detector \citep[HRC-I;][]{2000SPIE.4012...68M}. In Table\,\ref{tab:obs_xray} we list all X-ray observations.

%
\begin{table*}[ht]
\caption{X-ray observations of nova \nova.}
\label{tab:obs_xray}
\begin{center}
\begin{tabular}{lrrrrrr}\hline\hline \noalign{\smallskip}
	Telescope/Instrument$^a$ & ObsID & Exp. time$^b$ & Date$^c$ & $\Delta t^d$ & Count Rate$^e$ & L$_{0.2-10.0}$ $^e$\\
	 & & [ks] & [UT] & [d] & [ct s$^{-1}$] & [erg s$^{-1}$]\\ \hline \noalign{\smallskip}
	\swift XRT & 00031851001 & 2.6 & 2010-10-27.56 & 16.05 & $< 6.7$ \tpower{-3} & $< 3.1$ \tpower{37}\\
	\swift XRT & 00031851002 & 4.8 & 2010-11-03.04 & 22.53 & (1.1 $\pm$ 0.2) \tpower{-2} &  (5.1 $\pm$ 0.9) \tpower{37}\\
	\swift XRT & 00031861002 & 3.9 & 2010-11-06.20 & 25.69 & (4.3 $\pm$ 1.5) \tpower{-3} &  (2.0 $\pm$ 0.7) \tpower{37}\\
	\swift XRT & 00031861001 & 7.7 & 2010-11-07.06 & 26.55 & (1.9 $\pm$ 0.2) \tpower{-2} &  (8.6 $\pm$ 0.9) \tpower{37}\\
	\swift XRT & 00031861003 & 7.7 & 2010-11-09.53 & 29.02 & (8.4 $\pm$ 1.4) \tpower{-3} &  (3.9 $\pm$ 0.6) \tpower{37}\\
	\swift XRT & 00031861004 & 7.9 & 2010-11-11.01 & 30.50 & (5.5 $\pm$ 1.2) \tpower{-3} &  (2.5 $\pm$ 0.6) \tpower{37}\\
	\swift XRT & 00031861005 & 6.8 & 2010-11-13.09 & 32.58 & (5.3 $\pm$ 1.3) \tpower{-3} &  (2.4 $\pm$ 0.6) \tpower{37}\\
	\chandra HRC-I & 12110 & 20.0 & 2010-11-14.17 & 33.66 & (2.1 $\pm$ 0.1) \tpower{-2} &  (7.9 $\pm$ 0.4) \tpower{37}\\
	\swift XRT & 00031861006 & 6.8 & 2010-11-15.16 & 34.65 & (7.7 $\pm$ 1.3) \tpower{-3} &  (3.5 $\pm$ 0.6) \tpower{37}\\
	\swift XRT & 00031861007 & 8.2 & 2010-11-17.09 & 36.58 & (3.0 $\pm$ 0.9) \tpower{-3} &  (1.4 $\pm$ 0.4) \tpower{37}\\
	\swift XRT & 00031861008 & 6.8 & 2010-11-19.58 & 39.07 & (2.7 $\pm$ 0.9) \tpower{-3} &  (1.2 $\pm$ 0.4) \tpower{37}\\
	\swift XRT & 00031861009 & 8.2 & 2010-11-21.03 & 40.52 & (1.9 $\pm$ 0.9) \tpower{-3} &  (0.9 $\pm$ 0.4) \tpower{37}\\
	\swift XRT & 00031861010 & 7.9 & 2010-11-23.05 & 42.54 & $< 6.0$ \tpower{-3} &  $< 2.8$ \tpower{37}\\
	\chandra HRC-I & 12111 & 19.9 & 2010-11-23.18 & 42.67 & (4.9 $\pm$ 1.9) \tpower{-4} &  (1.8 $\pm$ 0.7) \tpower{36}\\
	\chandra HRC-I & 12112 & 19.9 & 2010-12-03.66 & 53.15 & (7.7 $\pm$ 2.4) \tpower{-4} &  (2.9 $\pm$ 0.9) \tpower{36}\\
	\chandra HRC-I & 12113 & 19.0 & 2010-12-12.56 & 62.05 & $< 2.7$ \tpower{-4} &  $< 1.0$ \tpower{36}\\
	\chandra HRC-I & 12114 & 20.0 & 2010-12-22.18 & 71.67 & $< 5.1$ \tpower{-4} &  $< 1.9$ \tpower{36}\\
	\xmm EPIC & 0650560201 & 18.7 & 2010-12-26.43 & 75.92 & (3.4 $\pm$ 0.8) \tpower{-3} &  (0.7 $\pm$ 0.2) \tpower{36}\\
	\xmm EPIC & 0650560301 & 21.6 & 2011-01-04.76 & 85.25 & (1.8 $\pm$ 0.7) \tpower{-3} &  (0.4 $\pm$ 0.2) \tpower{36}\\
	\xmm EPIC & 0650560401 & 11.9 & 2011-01-15.01 & 95.50 & (2.9 $\pm$ 0.8) \tpower{-3} &  (0.4 $\pm$ 0.2) \tpower{36}\\
	\xmm EPIC & 0650560501 & 6.2 & 2011-01-25.30 & 105.79 & $< 5.4$ \tpower{-3} &  $< 0.9$ \tpower{36}\\
	\xmm EPIC & 0650560601 & 16.2 & 2011-02-04.00 & 115.49 & (2.2 $\pm$ 0.8) \tpower{-3} &  (0.5 $\pm$ 0.2) \tpower{36}\\
	\chandra HRC-I & 13178 & 17.5 & 2011-02-17.15 & 128.64 & $< 2.9$ \tpower{-4} &  $< 1.1$ \tpower{36}\\
	\chandra HRC-I & 13179 & 17.5 & 2011-02-27.25 & 138.74 & $< 4.2$ \tpower{-4} &  $< 1.6$ \tpower{36}\\
	\chandra HRC-I & 13180 & 17.3 & 2011-03-10.12 & 149.61 & $< 6.4$ \tpower{-4} &  $< 2.4$ \tpower{36}\\
\hline
\end{tabular}
\end{center}
\noindent
Notes:\hspace{0.1cm} $^a $: Telescope and instrument used for observation; $^b $: Dead-time corrected exposure time of the observation; $^c $: Start date of the observation; $^d $: Time in days after the outburst of nova \nova in the optical on 2010-10-11.51 (MJD = 55480.51); $^e $: Source count rates, X-ray luminosities (unabsorbed, black body fit, 0.2 - 10.0 keV) and upper limits were estimated according to Sect.\,\ref{sec:results}.\\
\end{table*}

Data analysis for the \swift XRT was carried out using the source statistics (\texttt{sosta}) tool within the HEAsoft XIMAGE package (version 4.5.1.). This approach included corrections for detector exposure (exposure maps created using the XRT software task \texttt{xrtexpomap}) and the point spread function (PSF) of the source (XIMAGE command \texttt{psf}).

The \xmm data were analysed using the XMMSAS v11.0 software \citep[\xmm Science Analysis System;][]{2004ASPC..314..759G}\footnote{http://xmm.esac.esa.int/external/xmm\_data\_analysis/}. Our data analysis techniques differ from the standard processing and are described in detail in \citet{2010A&A...523A..89H}. The \chandra HRC-I observations were reduced with the CIAO v4.4 software package \citep[Chandra Interactive Analysis of Observations;][]{2006SPIE.6270E..60F}\footnote{http://cxc.harvard.edu/ciao/} and with adapted versions of XMMSAS tools, starting with a re-processing of the level 2 event files. With respect to \citet{2010A&A...523A..89H}, our \chandra data reduction procedures have been updated to allow for better treatment of the HRC-I PSF. We here used XMMSAS only to create a background map (tool: \texttt{esplinemap}) and based the source detection solely on the CIAO \texttt{wavdetect} algorithm. For an extensive description of our \chandra analysis pipeline see Hofmann et al. (in preparation). The astrometry for the detected X-ray sources was corrected with respect to the catalogue of \citet{2002ApJ...578..114K}, which was calibrated astrometrically using the Two Micron All Sky Survey \citep[2MASS,][]{2003tmc..book.....C}.

X-ray spectra, extracted from the \xmm observations, were analysed in XSPEC \citep[][version 12.7.0]{1996ASPC..101...17A}. For the resulting spectral model, specific energy conversion factors were estimated in XSPEC using the \texttt{fakeit} command. We searched for variability within the individual observations by extracting light curves using \texttt{evselect} for \xmm and \texttt{dmextract} for \chandra data. Additionally, \chandra detections were analysed using the \texttt{glvary} tool, which applies the algorithm of \citet{1992ApJ...398..146G} to classify source variability. A search for light curve periodicities was conducted using the XRONOS tasks of HEASARCs software package FTOOLS\footnote{http://heasarc.gsfc.nasa.gov/ftools/} \citep{1995ASPC...77..367B}.\\

\subsection{Optical observations}
\label{sec:obs_opt}
Motivated by the discovery of the new X-ray source in an \m31 GC, we re-analysed archival optical observations in search for a counterpart. The data set consisted of observations carried out with small telescopes participating in nova search projects and of \m31 monitoring observations obtained during the Pan-STARRS~1 survey (Panoramic Survey Telescope and Rapid Response System). This allowed for a detailed coverage of the \bol light curve.

\subsubsection{Observations with small telescopes}
\label{sec:opt_small}
Optical data were obtained at five different observatories in the context of regular \m31 monitoring programmes with the following telescopes: (a) the Livermore Optical Transient Imaging System \citep[\textit{Super-LOTIS},][]{2008AIPC.1000..535W}, a robotic 60 cm telescope with an E2V CCD (2kx2k) located at Steward Observatory, Kitt Peak, Arizona, USA (observer: G.G.~Williams); (b) a Meade 200R 40~cm f/9.8 reflector, plus SBIG STL1001E camera, at \textit{Miyaki-Argenteus} observatory, Japan (observers: F.~Kabashima and K.~Nishiyama); (c) a 50~cm f/6 telescope, with BITRAN BN-52E(KAF-1001E) camera, located at \textit{Itagaki} Astronomical Observatory, Japan (observer: K.~Itagaki); (d) a 35~cm f/7.5 Celestron C14 Schmidt-Cassegrain telescope at \textit{Xingming} observatory, China (observer: G.~Xing); (e) a 65~cm telescope, with G2CCD-3200 camera, at \textit{Ond\v{r}ejov} observatory, Czech Republic (observers: K.~Hornoch, M.~Wolf, P.~Hornochov\'a, P.~Ku\v{s}nir\'ak and P.~Zasche). While observatories (b)-(d) took unfiltered images, Super-LOTIS used a Johnson $R$ and Ond\v{r}ejov a Kron-Cousins $R$ filter. Observation dates are given in Table\,\ref{tab:obs_opt}.

Images from the first four telescopes were reduced and calibrated in a homogeneous way. This procedure made use of the TERAPIX software packages \textit{SExtractor} \citep[][]{1996A&AS..117..393B} for source extraction, \textit{SWarp} \citep[][]{2002ASPC..281..228B} for image stacking, and \textit{SCAMP} \citep[][]{2006ASPC..351..112B} for image calibration. Image reduction procedures corrected for the strong background light of \m31 and specific detection thresholds were used to create clean source catalogues. The astrometric and photometric solutions were computed in SCAMP using $R$ magnitudes from the \m31 part of the Local Group Galaxy Survey \citep[LGGS,][]{2006AJ....131.2478M}. Photometric uncertainties were estimated from all sufficiently star-like objects in a $1$ mag range around the magnitude of \bolk.  

Data from the Ond\v{r}ejov Observatory were analysed using the SIMS\footnote{http://ccd.mii.cz/} and Munipack\footnote{http://munipack.astronomy.cz/} programmes. Reduced images of the same series were co-added to improve the signal-to-noise ratio (total exposure time varied from 600~s up to 1800~s). The gradient of the galaxy background of co-added images was flattened by the spatial median filter using SIMS. These processed images were used for aperture photometry, carried out in GAIA\footnote{http://www.starlink.rl.ac.uk/gaia}. Relative photometry was performed using brighter field stars that were calibrated using standard Landolt fields.

The resulting magnitudes of \bol from both samples agree well (see Table\,\ref{tab:obs_opt} and Sect.\,\ref{sec:results}).

\subsubsection{Observations with Pan-STARRS~1}
\label{sec:opt_pan}
The PAndromeda survey was designed to identify gravitational microlensing events towards \m31 within the Pan-STARRs survey (PS1). It monitors \m31 for five months per year and 30 minutes per night (including overhead). With the 7 deg$^2$ field of the Giga Pixel Camera (GPC) mounted on the 1.8~m telescope on Haleakala (Maui, US) the entirety of \m31 can be observed with one pointing. 

The data taken in the first PAndromeda season as well as the data reduction are described in \citet{2012AJ....143...89L}. In short, the astrometric accuracy of the final data is of the order of $0\farcs1$, when compared to the SDSS-DR7 catalogue which was not used to derive the astrometric solution. The photometric accuracy can be seen in figure 10 of \citet{2012AJ....143...89L}; it is 0.01 mag for a magnitude of 16 in the bulge of M31. Since the colour terms of the $r_{p1}$ and $i_{p1}$ Pan-STARRS filters relative to the corresponding SDSS filter systems are small, $(r_{P1}-r_{SDSS})=0.000-0.007(g-r)_{SDSS}$ and $(i_{P1}-i_{SDSS})=0.004-0.014(g-r)_{SDSS}$, the Pan-STARRS magnitudes can be considered as equivalent to the SDSS magnitudes for this work.

Light curves for variable objects are derived by difference imaging technique \citep{1998ApJ...503..325A} using the implementation of \citet{2002A&A...381.1095G,2004ASPC..314..456G} as described in \citet{2012AJ....143...89L}. The PAndromeda data were successfully searched for microlensing events \citep{2012AJ....143...89L} and are currently being analysed to study cepheids, eclipsing binaries and novae. The brightest and shortest nova found in PAndromeda data up to now was the one discussed in this paper. Its light curves in the $r_{p1}$ and $i_{p1}$ filters are given in Table\,\ref{tab:obs_pan2} (in AB-magnitudes).

\section{Results}
\label{sec:results}
%
The position of the new X-ray source was determined from \chandra observation 12110, because the HRC-I detector has the best spatial resolution and the source had the highest count rate in this pointing (see Table\,\ref{tab:obs_xray}). The coordinates are RA = 00h42m43.70s, Dec = +41$\degr$12$\arcmin 43\,\farcs0$ (J2000, $1\sigma$ accuracy of $0\,\farcs8$), which agree well (distance $0\,\farcs4$) with the position of the \m31 GC \bolk: RA = 00h42m43.681s, Dec = +41$\degr$12$\arcmin 42\,\farcs70$ according to the Revised Bologna Catalogue of \m31 globular clusters and candidates \citep[RBC; version 4.0, Dec 2009;][]{2004A&A...416..917G}. No X-ray source was previously known at this position \citep[e.g.][]{2002ApJ...578..114K,2002ApJ...577..738K,2004ApJ...609..735W,2005A&A...434..483P,2011A&A...534A..55S}.

\begin{figure}[t!]
	\resizebox{\hsize}{!}{\includegraphics[angle=90]{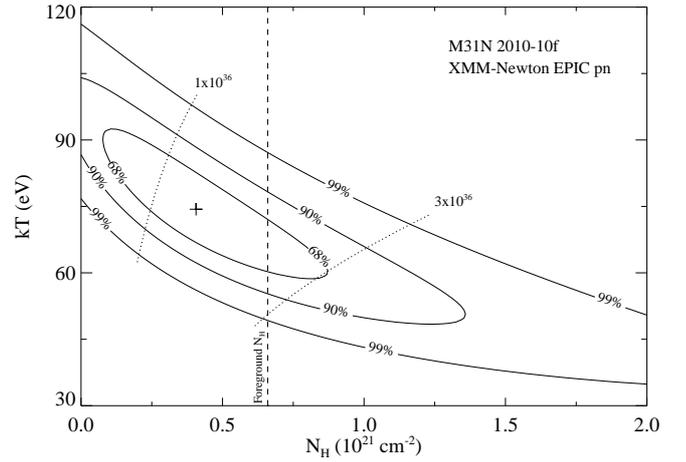}}
	\caption{Column density (\nh) - temperature (kT) contours inferred from the simultaneous black body fit to the \xmm EPIC pn spectra of \novak. Indicated are the formal best-fit parameters (cross), the lines of constant X-ray luminosity (0.2-10.0 keV, dotted lines), and the Galactic foreground absorption (dashed line).}
	\label{fig:xray_spec}
\end{figure}

We simultaneously fitted the \xmm EPIC pn spectra (single-pixel events) of the X-ray source using an absorbed black body model with best-fit parameters $kT = 74^{+30}_{-26}$ eV and \nh = ($0.4^{+0.9}_{-0.4}$) \hcm{21}, resulting in a formal unabsorbed luminosity of 1.3 \ergs{36}. Therefore, the source can be classified as an SSS. Confidence contours for absorption column density and black-body temperature are shown in Fig.\,\ref{fig:xray_spec}. The uncertainty ranges (90\% confidence) for the black-body parameters are relatively large, because the luminosity of the source had already declined significantly by the time of the \xmm observations (see Table\,\ref{tab:obs_xray}). Unfortunately, the \chandra HRC-I detector has no energy resolution to derive a spectrum from the initial, more luminous detections. Although the source was also bright during the \swift pointings, the shorter exposure time and smaller effective area of the XRT led to a combined spectrum in agreement with the \xmm spectrum, but with larger uncertainties. 

High-resolution spectra of Galactic novae \citep[e.g.][]{2008ApJ...673.1067N,2011ApJ...733...70N,2012ApJ...745...43N} clearly show a variety of absorption and emission features, underlining the fact that black body fits merely provide a qualitative parametrisation of SSS spectra and not a physically realistic model \citep[see also][]{1991A&A...246L..17G,1997ARA&A..35...69K}. For individual nova SSS spectra, results based on assuming black body models have to be interpreted with great care, but general population trends appear to be describable by black-body temperatures \citep{2011A&A...533A..52H}.

Based on the black body model, we derived energy conversion factors for the different X-ray detectors in XSPEC, which were used to compute unabsorbed fluxes from the instrumental count rates listed in Table\,\ref{tab:obs_xray}. The unabsorbed luminosities, which are given in Table\,\ref{tab:obs_xray}, assume an \m31 distance of 780~kpc \citep[][]{1998AJ....115.1916H,1998ApJ...503L.131S}. No significant short-term variability was found in any of the individual observations.

%
\begin{figure*}[t!]
	\resizebox{\hsize}{!}{\includegraphics[angle=0]{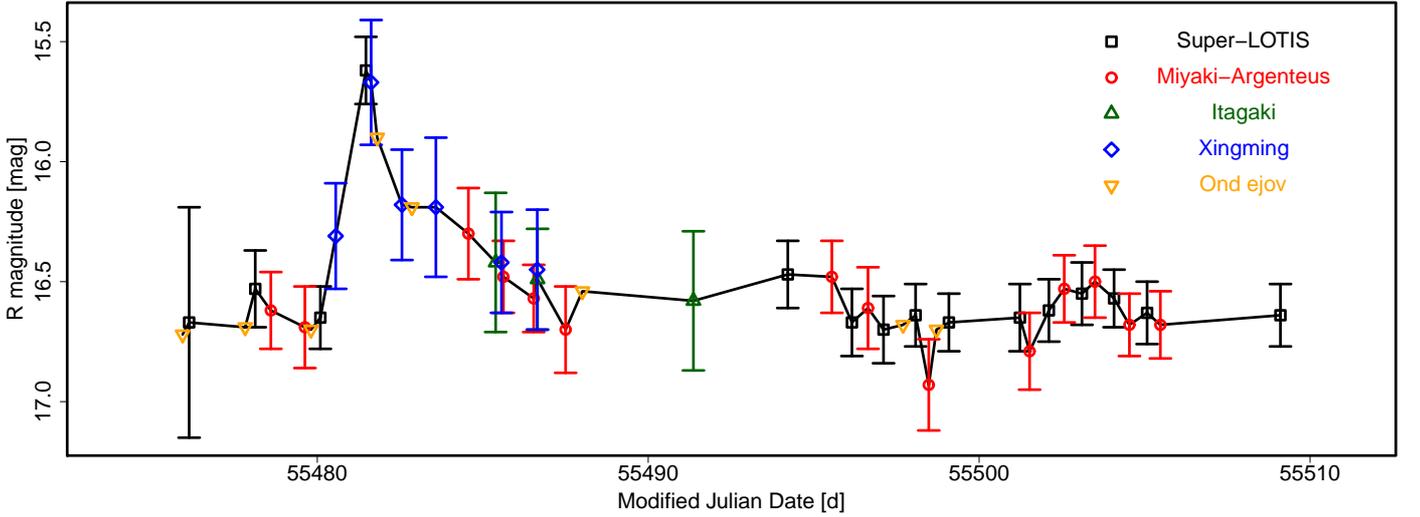}}
	\caption{Optical $R$ band light curve of \bol consisting of data from five observatories indicated by different colours and symbols. One sigma confidence ranges are indicated by error bars. For the Ond\v{r}ejov data the error ranges are about the size of the symbols.}
	\label{fig:lc_opt}
\end{figure*}

As mentioned above, the discovery of the third SSS in an \m31 GC motivated a search for an optical counterpart, because one of the two other sources was identified with the first nova in the \m31 GC system (\sqk, \mgck). Archival optical data from Super-LOTIS showed \bol at a constant $R = 16.7\pm0.1$ mag during the time from 2007 until the beginning of October 2010 and from November 2010 onward. During October 2010, the brightness of the GC experienced a significant increase by about one magnitude (see Table\,\ref{tab:obs_opt}). Figure\,\ref{fig:lc_opt} shows that observations from five different telescopes indicate a potential nova outburst. This nova candidate is hereafter called \novak, following the naming convention described in \citet{2007A&A...465..375P}. 

\begin{figure}[t!]
	\resizebox{\hsize}{!}{\includegraphics[angle=0]{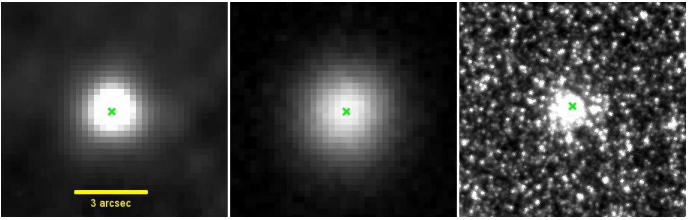}}
	\caption{Astrometric agreement of \nova (green cross) and the GC \bol in a PS1 image (left) and a HST image (right) from ACS 814 nm archive data of HST Cycle 18 proposal 12058 by Dalcanton, J. The middle image shows the PS1 difference frame in $r_{P1}$ at the peak of the nova. The GC was subtracted in this frame and the image was used to compute the position of the nova as indicated by the green crosses in this figure. All three images have the same size (FOV $9\arcsec \times 9\arcsec$).}
	\label{fig:nova_pos}
\end{figure}

\begin{figure}[t!]
	\resizebox{\hsize}{!}{\includegraphics[angle=0]{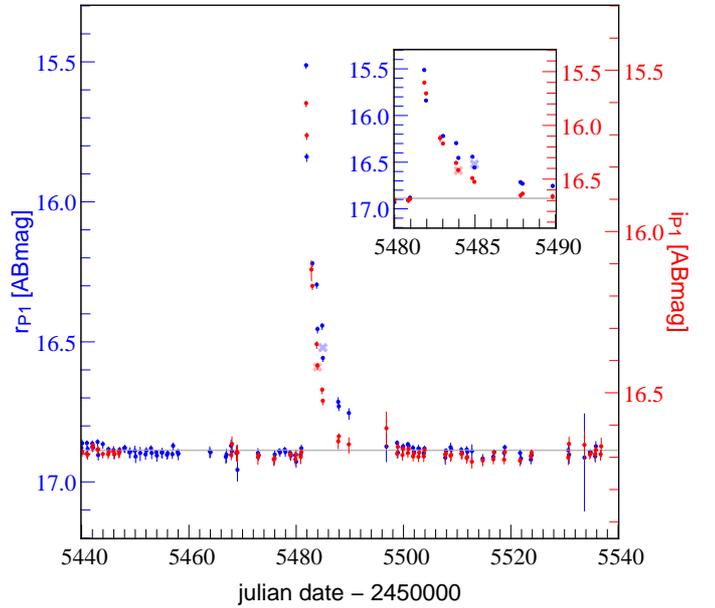}}
	\caption{Light curve of nova \nova as measured from the PAndromeda data. The $r_{P1}$ and $i_{P1}$-band AB-magnitudes and their errors are shown as blue and red data points and error bars. The measurements include the brightness of the globular cluster ($r_{P1}=16.89$ and $i_{P1}=16.68$ mag). The two large crosses in light blue and red show the estimated brightness and point in time $t_2$ where the nova faded by 2 magnitudes.}
	\label{fig:lc_pan}
\end{figure}

In Fig.\,\ref{fig:nova_pos} we illustrate the good agreement between the position of \novak, as inferred from the PS1 difference image, and the GC \bolk. Light curves (in $r_{p1}$ and $i_{p1}$ AB-magnitudes) of \nova from PAndromeda data are shown in Fig.\,\ref{fig:lc_pan}.

%
%

%

\begin{figure*}[t!]
	\resizebox{\hsize}{!}{\includegraphics[angle=0]{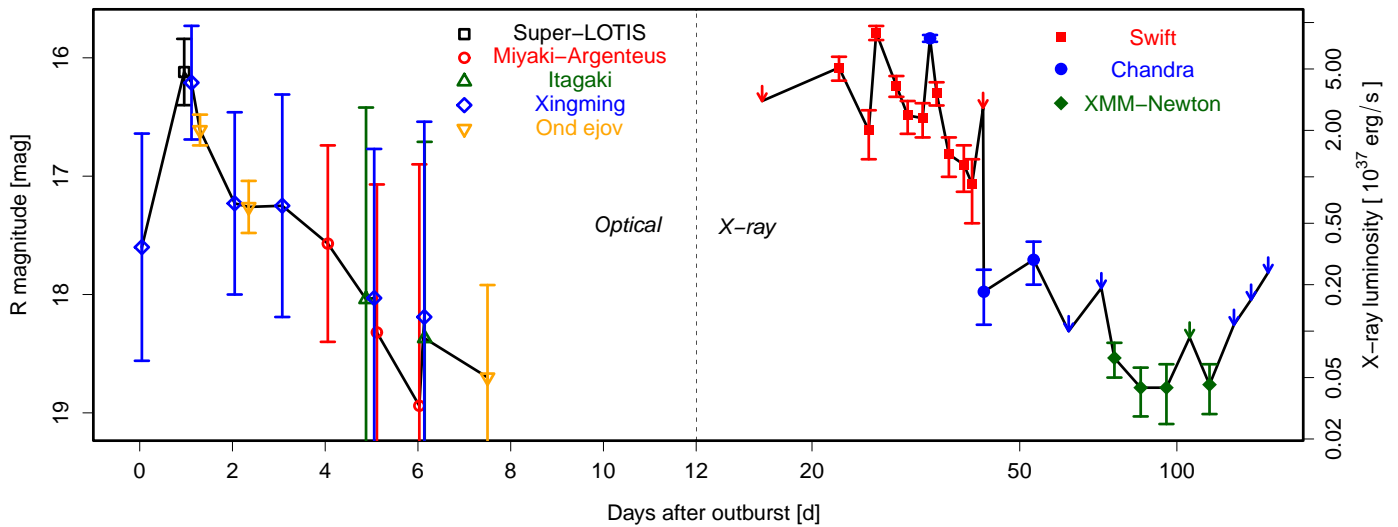}}
	\caption{Light curve of nova \nova in the optical $R$ band (left part, nova magnitude assuming \bol having constant $R = 16.7$ mag) and in the X-ray range (right part), assuming a nova outburst on MJD = 55480.32 (see text). Note the logarithmic luminosity scale for the X-ray light curve and that the scale of the ordinate changes from linear to logarithmic at day 12. Error bars show $1 \sigma$ confidence range, down-pointing arrows indicate $3 \sigma$ upper limits. Data from different observatories are colour- and symbol-coded.}
	\label{fig:lc_nova}
\end{figure*}

Assuming the quiescence magnitude of \bol as given above, we computed the magnitudes of \nova using the standard formula. The results are given in Table\,\ref{tab:obs_opt} for the duration of the outburst and illustrated in Fig.\,\ref{fig:lc_nova} together with the X-ray light curve. Using our high-cadence observational coverage, we could determine the time of the nova outburst with high precision to MJD = $(55480.51\pm0.05)$~d. This assumes that the first Xingming observation in Fig.\,\ref{fig:lc_nova} detected the beginning of the outburst, whereas the earlier PAndromeda observation (see Fig.\,\ref{fig:lc_pan}) saw \bol still at quiescence. The observed peak magnitudes of \nova are $R = (16.1\pm0.3)$ mag, detected in a Super-LOTIS observation on MJD = $55481.47$ (see Table\,\ref{tab:obs_opt}), as well as $r = (15.512\pm0.009)$ mag and $i = (15.602\pm0.009)$ mag from slightly earlier PAndromeda observations on MJD = $55481.34$ and $55481.35$ (see Table\,\ref{tab:obs_pan2}), respectively. 

The PAndromeda light curve provides the data points closest to the outburst maximum of the nova and therefore these data are most suited to determine $t_2$, which is the time (in days) for the nova to decline by 2 magnitudes:

\begin{equation}
 m_{\mbox{\tiny{t2}}} = -2.5 \log((F_{\mbox{\tiny{GC}}}+10^{-0.4\times 2\mbox{\tiny{mag}}} (F_{\mbox{\tiny{max}}}-F_{\mbox{\tiny{GC}}}))/3631 \mbox{\small{Jy}}) \; . \\
\end{equation}

Using those points in the light curve that have the most similar brightness to $m_{t2}$ gave upper limits of $t_{2,R}\le 3.12 d$ and $t_{2,I}\le2.12 d$, making \nova a very fast nova in the classification system of \citet{1964gano.book.....P}. The brightness of the cluster was determined by aperture photometry to be $F_{\mbox{\tiny{R,GC}}} = 63.805\times10^{-5} Jy$ and $F_{\mbox{\tiny{I,GC}}} = 77.462\times10^{-5} Jy$ (16.89 mag and 16.68 mag).

%
With respect to the derived outburst date, the X-ray time scales of the nova can be estimated as follows: SSS turn-on time $\eton = (19\pm3)$~d and turn-off time $\etoff = (40\pm1)$~d. While \ton is clearly constrained (see Table\,\ref{tab:obs_xray}), \toff is not as straightforward to determine. Between the second-last \swift observation and the second \chandra observation the luminosity of the SSS dropped by an order of magnitude (see Table\,\ref{tab:obs_xray} and Fig.\,\ref{fig:lc_nova}). We interpret this observation as the end of the constant bolometric luminosity phase \citep{1985ApJ...294..263M,2005A&A...439.1061S}, which indicates the cessation of stable hydrogen burning \citep{2010ApJ...709..680H}. 

Following our initial discovery alert \citep{2010ATel.3019....1H}, \citet{2010ATel.3074....1S} carried out optical spectroscopy to confirm the nova candidate. However, in their observation about 32~d after outburst they found no obvious Balmer emission lines that would indicate a recent nova. Such a short visibility in particular in H$\alpha$ is unusual for an \m31 nova and might appear troubling a first. However, in the following we will outline how this fits into the picture that has emerged for \novak.

\nova was a very rapidly evolving object, much more so than the first \m31 GC nova (M31N~2007-06b in Bol~111), which indeed was found to show strong H$\alpha$ emission for at least five weeks post-discovery (\sqk). But even compared to the relatively fast M31N~2007-06b, the X-ray time scales for \nova were significantly shorter: SSS turn-on time $(19\pm3)$~d vs $(87\pm54)$~d and turn-off time $(40\pm1)$~d vs $(452\pm57)$~d. There is only an upper limit of $t_{2,R}<18$~d known for M31N~2007-06b to compare to the very fast $t_{2,R}\le 3.12 d$ of \novak, but in \citet{2011A&A...533A..52H} we found that short turn-on times for \m31 novae were correlated with fast expansion velocities of the ejected envelope as well as with rapid optical declines. \citet{1990ApJ...356..472C} noted that after the maximum H$\alpha$ light of a nova (which occurs after the optical peak) the decline in H$\alpha$ matched the decline in broad band B magnitude. Because our R band includes the H$\alpha$ line, we can assume that \nova would have experienced an $t_{2,H\alpha} \lesssim 3$~d. Although \citet{2010ATel.3074....1S} give no detection threshold for their observations, we assume that even from a bright H$\alpha$ peak magnitude \nova could have faded sufficiently fast to be not detectable anymore in their measurements.

Moreover, according to current understanding \citep[e.g.][]{2006ApJS..167...59H} the delay of the SSS turn-on with respect to the optical maximum provides an indirect hint for the presence of an expanding envelope (which earlier H$\alpha$ observations should have detected) that at a certain time becomes optically thin to soft X-rays. Therefore, even without a confirmation and classification based on H$\alpha$ detections or optical spectra, all evidence strongly indicates a nova outburst. The signature of the optical transient fits the amplitude as well as the shape of a nova outburst and the X-ray spectrum points towards a nuclear burning WD. There is no other type of object known that has these observable properties. Consequenly, we interpret our observations as revealing the outburst of a nova in the GC \bolk.

Although there is an irregular variable present in the Wendelstein Calar Alto Pixellensing Project (WeCAPP) catalogue of \citet{2006A&A...445..423F} at a position only $0\farcs35$ away from \bolk, a quick inspection of the corresponding light curve did not reveal a previous nova outburst. Similarly, no signatures of additional CNe were found in the WeCAPP light curves of GCs from the catalogues of \citet{2004A&A...416..917G} and \citet{2010MNRAS.402..803P} (see also Sec.\,\ref{sec:disc_rate_opt}).

An additional optical light curve of \bol is given by \citet{2012ApJ...752..133C} based on Palomar Transient Factory data. These authors confirm the outburst and their light curve plots agree with our classification of \nova as a very fast nova.

\section{Discussion}
\label{sec:discuss}
%
\subsection{Properties of nova \nova and \m31 GC novae}
\label{sec:disc_nova}
\nova is only the second confirmed nova in the \m31 GC system to date. Its observational properties in the optical (relatively bright maximum magnitude, fast decline) and X-rays (short SSS duration, high black-body temperature) coherently point towards an underlying massive WD \citep[see e.g.][]{1992ApJ...393..516L,2005A&A...439.1061S,2006ApJS..167...59H,2010ApJ...709..680H}. By comparing the models of \citet{2006ApJS..167...59H}, we estimated \mwd~$\gtrsim 1.3$~\msun. The relations between the measured properties $kT$, \ton, \toff and $t_{2,R}$ do not deviate significantly from the general population trends presented in \citet{2011A&A...533A..52H}.

Since binary systems with massive WDs are expected to be found in stellar populations younger than those of GCs \citep[e.g.][]{1998ApJ...506..818D,2011A&A...533A..52H} the properties of \nova are a noteworthy finding. It becomes even more remarkable, because the first \m31 GC nova, M31N~2007-06b in Bol~111, displayed features normally associated with young stellar populations (He/N spectrum and broad Balmer emission lines, see \sqk) and the X-ray properties of a third, putative GC nova (in Bol~194) reported in \mgc also indicate a massive WD (short SSS duration, high black-body temperature). All three GCs are old systems (age $> 3$~Gyr) with low metallicities of [Fe/H] $< -1.0$ \citep[see e.g.][and references therein]{2011AJ....142....8S}. 

The trend within GC novae towards hot, short SSS stages is unlikely to be caused by observational selection effects. In the field, novae with massive WDs are dominating the \textit{observed} mass distribution because of their short recurrence times \citep[e.g.][]{1986ApJ...308..721T}. However, those novae have also much shorter SSS durations and require high-cadence monitorings like our programme to find them. On the other hand, novae with low-mass WDs are visible in X-rays for years, some even for a decade \citep[see][]{2011A&A...533A..52H}. Those objects would have been detected in the combined X-ray data from extensive monitoring of the \m31 central area \citep{2010A&A...523A..89H,2011A&A...533A..52H}.
In the optical, slow novae are considerably more difficult to detect against the GC background, as they are optically fainter.

The non-detection of slow novae in \m31 GCs together with the similarities between the known novae might therefore present a challenge to the current understanding of nova populations. This underlines the importance and wide discovery space of a regular monitoring of our neighbour galaxy with X-ray telescopes, in particular if undertaken with high-cadence observations.

A possible answer to why only fast novae have been discovered in \m31 GCs so far might be found in a recent suggestion by \citet{2012MNRAS.423....2M} on how to enhance helium abundances in GCs. They discuss the impact on GC abundances of He-rich ejecta from novae powered by accretion of the intra-cluster medium (ICM) onto massive WDs \citep[see also][]{2011ApJ...735...25N}. The scenario of nova outbursts powered by accretion of interstellar matter was already mentioned by \citet{2005ApJ...629..750D} while examining type Ia supernovae (SNe~Ia) rates in radio galaxies. According to the estimate of \citet{2012MNRAS.423....2M}, there should be a large number of nova outbursts in GCs due to ICM accretion, but many of them will probably be obscured by the same high-density ICM from which they are accreting. Interestingly, this scenario should favour (massive) ONe WDs, as such systems form first, and could therefore explain the observations.

\subsection{The nova rate in \m31 GCs from the optical WeCAPP survey}
\label{sec:disc_rate_opt}
To determine a nova rate in \m31 GCs from the WeCAPP optical survey data, we used as reference the catalogue of \citet{2010MNRAS.402..803P}. From a total of 572 confirmed GCs in the catalogue (416 old and 156 young GCs), 80 overlap with the WeCAPP field (RA = 00h41m58.1s to 00h43m29.4s, Dec = +41$\degr$07$\arcmin 43\arcsec$ to +41$\degr$24$\arcmin 23\arcsec$; J2000.). For 78 GCs we were able to derive light curves in the $R$~band in our WeCAPP data, whereas two objects are too close to a saturated star to obtain reliable measurements. The coordinates given in \citet{2010MNRAS.402..803P} agree within an accuracy of less than $0\farcs5$ with the centroids on the source positions of the WeCAPP frames \citep[see][]{2012A&A...537A..43L}.

We carried out a search for additional nova outbursts, according to the criteria described in \citet{2012A&A...537A..43L}, on the 78 remaining GC positions. We applied a slightly modified asymmetry criterion using $10\sigma$ outliers and $a>6$ (an empirical asymmetry parameter describing the balance between outliers in both directions) only for light curves that had at least seven upper outliers. With these criteria, no new nova was detected in the WeCAPP light curves of the 78 GCs, whereas all 91 WeCAPP novae and \nova were found.

To determine a detection efficiency we performed Monte Carlo simulations using our 91 WeCAPP novae and the \nova light curve as a sample and interpolating linearly between the magnitude values of the data points. For all 78 individual GCs each of the 92 nova light curves was simulated \power{4} times, where we equally distributed the maximum time of the different WeCAPP novae over the WeCAPP survey time between Julian dates of 2450685.5 and 2454535.3.

\begin{figure}[t!]
	\resizebox{\hsize}{!}{\includegraphics[angle=0]{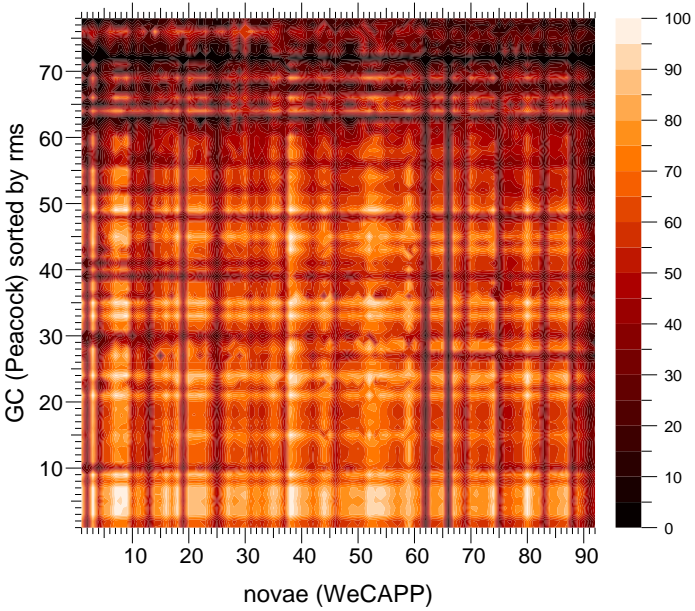}}
	\caption{Detection efficiency in percent (see scale) depending on 78 individual GC and 92 different novae types. The GCs, from the catalogue of \citet{2010MNRAS.402..803P}, are sorted by increasing rms of their light curves. The light curve of \nova is number 92.}
	\label{fig:deteff_wecapp}
\end{figure}

The results of the simulation are illustrated in Fig.\,\ref{fig:deteff_wecapp}, which shows that the detection efficiency mainly depends on the individual nova light curve (vertical stripes) and the rms of the GC light curves (horizontal stripes). Nova light curves can differ strongly in maximum brightness and duration, i.e. the detection window varies. For the GC light curves Fig.\,\ref{fig:deteff_wecapp} shows that the detection efficiency tends to decrease with increasing rms. Deviations from this trend in individual GCs are caused by the quite diverse time sampling and different noise quality of their light curves. In particular, some GCs show an intrinsic variability that reduces nova detections with our criteria. Therefore the measurable nova rate can differ, depending on the type of novae present in the particular GC, and Monte Carlo simulations were necessary to estimate the detection efficiency.

%

For nova light curves similar to that of \nova the mean detection efficiency is 27\%. With only one detected nova in 78 GCs during the WeCAPP survey period of 10.54 yr, this results in a tentative nova rate of 0.0045 novae yr$^{-1}$ GC$^{-1}$. The corresponding 95\% confidence Poisson upper limit is 0.025 novae yr$^{-1}$ GC$^{-1}$. Surveying all confirmed GCs of \citet{2010MNRAS.402..803P} for one year, we would expect two to three novae to be detected. For all nova light curves and all confirmed GCs, the mean detection efficiency becomes 51\%, which reduces the above rate estimate and confidence limit by a factor of about two. For the final data release of Pan-STARRS 1 with its key project PAndromeda we expect to further constrain these numbers.

\subsection{The nova rate in \m31 GCs from the X-ray monitoring}
\label{sec:disc_rate}
Assuming that both SSSs presented in \mgc were actually novae, we here estimate the nova rate for the \m31 GC system based on X-ray observations. In five recent X-ray monitoring campaigns aimed at the \m31 central region we detected three GC novae. The campaigns considered here were carried out with a 10~d cadence and are summarised in \citet{2011A&A...533A..52H} (2007/8 and 2008/9) and Henze et al. (in preparation) (2009/10, 2010/11, and 2011/12). The first dedicated monitoring campaign for X-ray emission from \m31 novae \citep[reported in][]{2010A&A...523A..89H} had a lower cadence of about 30~d and is therefore not included in our estimate. Since the two sources reported in \mgc were only found at large off-axis angles in the \chandra HRC-I observations of the monitoring (the \xmm EPIC field of view is slightly smaller), only HRC-I pointings are considered here. The \chandra part of each individual campaign covered a time span of one to two months, resulting in a total effective survey time of nine months.

Recently, \citet{2011A&A...533A..52H} compiled a catalogue of 60 novae with an X-ray counterpart in \m31. Even the fastest of these objects generally had an SSS duration of $\sim 20-30$~d. Therefore, we assume that a 10~d cadence monitoring should be able to detect all SSS counterparts of hypothetical GC novae during the time of the coverage. This means that in contrast to our estimate in \mgck, the SSS duration of the nova is not a critical parameter.

Of the approximately 650 confirmed GCs known in \m31 today \citep[see the RBC;][]{2004A&A...416..917G}, 160 are located within the field of view of the \chandra HRC-I (assuming a $16\arcmin$ radius around the \m31 centre). Of those, we expect about 50\% to be located within or behind the \m31 disk, as current GC catalogues appear to be complete except for a few objects located behind dust lanes \citep{2011AJ....141...61C}. This assumption is supported by a comparison of GC reddening estimates from the work of \citet{2008MNRAS.385.1973F} with an \m31 reddening map derived by \citet{2009A&A...507..283M}, for which we found approximately half of the GCs to have higher extinctions than the \m31 disk around their position \citep[see also][]{2004ApJ...616..821T}.

Our nova rate estimate needs to take into account the higher extinction for GCs in and particularly behind the \m31 disk, as supersoft X-rays are strongly attenuated by a high foreground column density. Additional absorption takes place within the intra-cluster medium and in the matter ejected by the current or previous nova outbursts. These latter effects are difficult to quantify. Here, we assume that the combined extinction for GC novae reduces the sample of GCs in which an SSS could be found in our monitoring by $\sim 50\%$. Therefore, the detection of three novae in about 80 GCs within nine months leads to a rate of $\sim0.05$ novae yr$^{-1}$ GC$^{-1}$.

Another effect that might influence whether the SSS state of a potential nova is detected or not, is the presence of other X-ray sources in the GC. About 40 confirmed GCs from the RBC are detected in X-rays within the \chandra HRC field of view. This number is slightly reduced for the smaller \xmm EPIC field. We constructed long-term X-ray light curves for \xmm and \chandra for each GC. In none of these light curves we did find unambiguous evidence for an additional outburst component. In case of \xmmk, using the spectral resolution of the EPIC detectors, we furthermore studied the evolution of hardness ratios (X-ray colours) over time, searching for the signature of a supersoft, transient source. No such event could be identified.

However, of the 40 HRC sources, four have average luminosities exceeding \oergs{38} and show intrinsic variability \citep[they are mostly black hole binary candidates, see e.g.][]{2012arXiv1203.2583B}. These properties would make it difficult to detect the presence of a fainter nova with a luminosity of only a few \oergs{37}. Although the hardness ratios of these sources are relatively stable over time, and no hint of a supersoft excess could be seen, we cannot exclude that a nova outburst could have been hidden in the existing variability. In the present context, the small number of four sources does not change the above estimate of about 80 GCs, in which novae could have been found, and therefore has no impact on our detection rate estimate of $\sim0.05$ novae yr$^{-1}$ GC$^{-1}$

Taken at face value, the nova rate derived from our X-ray monitoring exceeds the optical estimate based on the WeCAPP data (see Sect.\,\ref{sec:disc_rate_opt}) by about one order of magnitude. This demonstrates the advantages of X-ray surveys over optical observations when searching for novae in GCs. However, that nova M31N 2007-06b (\sqk) probably was only missed by the WeCAPP project because it was outside the field of view. The X-ray rate is also higher by a factor of 10 or 25 compared to upper limits from earlier optical surveys \citep[e.g. 0.005 yr$^{-1}$ GC$^{-1}$ in][]{1992BAAS...24.1237T} or simple estimates based on nova rates in elliptical galaxies (0.002, see \mgc and references therein), respectively.

Applying Poisson statistics, our X-ray discoveries (lower 95\% confidence limit: 0.01 novae yr$^{-1}$ GC$^{-1}$) are still consistent within the 95\% confidence limits with the WeCAPP estimate. However, this does not agree with the other two estimates within the 95\% confidence range (which corresponds to 0.8-11.7 novae yr$^{-1}$ for the entire GC system). 

Even without taking into account any extinction effects, our results remain significant. Considering all 160 GCs in our field of view still produces a (factor two lower) nova rate that is higher by a factor of 5 or 12 compared to the optical upper limits for \m31 GCs and elliptical galaxy rates and excludes those estimates on the 95\% confidence level.

For a conservative estimate, we assume that by chance the entire \m31 GC nova production took place in only $80/650 \sim 12\%$ of GCs, during the time of our monitoring. This approach leads to a rate of 0.006 novae yr$^{-1}$ GC$^{-1}$ and a 95\% confidence lower limit of 0.001 novae yr$^{-1}$ GC$^{-1}$. These numbers are comparable to the upper limits from optical surveys and the rates in ellipticals. However, the probability for this scenario is only 0.2\%.

\subsection{Expected nova rates in \m31 GCs and the overabundance of binary systems}
\label{sec:disc_gc_field}
Now we estimate the expected nova rate for the \m31 GC system assuming that in GCs CNe are produced with the same efficiency as in the field. For this, we compared the K~band magnitudes of \m31 and its GC system and used this relation to scale the overall \m31 nova rate. The RBC includes infrared measurements \citep[from the 2MASS project;][]{2006AJ....131.1163S} for about 350 confirmed GCs with a total magnitude of K $\sim7.1$~mag. We assume that the remaining about 300 confirmed GCs were too faint to be detected by 2MASS \citep[detection limit K~$\sim15.3$~mag;][]{2003tmc..book.....C}, because they show (a) a similar spatial distribution as the GCs with K~band magnitude and (b) a significantly fainter average magnitude in the R and I bands. Adding 300 objects with magnitudes of K~$= 16-17$~mag only changes the total magnitude to K $\sim7.0$~mag, which we used as a conservative estimate for all confirmed \m31 GCs. For an \m31 total K~band magnitude of 0.98 \citep[2MASS;][]{2006AJ....131.1163S}, this means that \m31 is about a factor of 250 brighter (in the K~band) than its GC system.

We scaled the total \m31 nova rate of about 65 yr$^{-1}$ \citep[][]{2006MNRAS.369..257D} with the luminosity to $65/250 = 0.26$ novae yr$^{-1}$ for the whole GC system. Assuming that the \m31 nova rate is dominated by an old (bulge) stellar population \citep[e.g.][]{1987ApJ...318..520C,1989AJ.....97.1622C}, the extrapolation from a total rate to a rate in the old populations of GCs is justified. Under the assumptions of Sect.\,\ref{sec:disc_rate} the observed rate is three novae in nine months of effective survey time or four novae yr$^{-1}$. Therefore, the derived GC nova rate might be higher than measured for the host galaxy by a factor of $4/0.26\sim15$, with the 95\% confidence lower limit allowing for an enhancement by at least a factor of $0.8/0.26 \sim 3$.

The above estimates indicate that an additional binary-forming mechanism might exist in GCs to increase the nova rate. One alternative scenario, increased accretion onto compact objects in GCs, is discussed in Sect.\,\ref{sec:disc_nova}. It is long known that low-mass X-ray binaries (LMXBs) are significantly overabundant in \m31 GCs \citep[e.g.][]{2005PASP..117.1236F}, which was explained by tidal captures of main-sequence stars by neutron stars \citep[][]{1975ApJ...199L.143C,1975MNRAS.172P..15F}. These processes are also expected to increase the number of WD binaries \citep{1983Natur.301..587H}. Indeed, Galactic GCs have been found to harbour a large number of cataclysmic variables (CVs), the majority of which are strongly suspected to have been formed dynamically \citep[e.g.][]{2007A&G....48e..12M,2011arXiv1112.1074K}.

Despite the recent progress in finding novae in GCs, made possible by the regular X-ray monitoring of \m31, the number of these objects is still small. Finding them is important for various reasons: (i) knowing the nova rate in GCs allows for a more accurate estimate of the total nova rate in a galaxy; (ii) comparing sample properties for field- and GC novae might provide additional information on CV formation processes; (iii) novae could play an important role in the ecosystem of the GC, such as removing part of the intra-cluster medium \citep{2011ApJ...728...81M}, enhancing its helium abundance \citep{2012MNRAS.423....2M} or creating unusual emission spectra \citep{2012MNRAS.423.1144R}; (iv) as potential progenitors of SNe~Ia novae could help to access the SN~Ia rate in GCs \citep[e.g.][]{2012A&A...539A..77V}.

Finally, a recent result by \citet{2010MNRAS.402..803P} is worth to be underlined. These authors reported a possible indication for a nova outburst in the \m31 GC Bol~383 in August 1991. At this time, observations by \citet{1992AJ....103..824R} showed Bol~383 about 0.5 mag brighter than measured before and afterwards. \citet{2010MNRAS.402..803P} concluded that a transient with $M_V \sim 8$ and a bluer colour index than the GC would have been needed to explain this phenomenon. No X-ray source is known in this GC. This cluster is also an old one \citep[][]{2011AJ....142....8S}. We encourage optical observers to check their archives, not only for additional observations of Bol~383 during the time of August 1991, but for possible indications of nova outbursts in GCs in general.

\section{Summary}
\label{sec:summary}
%
We presented the discovery and properties of the second confirmed CN in a \m31 GC based on high-cadence optical and X-ray monitoring observations. This object, named \novak, exhibits the characteristics of harbouring a massive WD (\mwd~$\gtrsim 1.3$~\msun) in its X-ray (high black-body temperature, short SSS phase) as well as in the optical properties (relatively bright peak magnitude, fast decline). Together with two additional GC novae (one suspected) from earlier work (\mgck), \nova allowed us to estimate the \m31 GC nova rate based on our regular X-ray monitoring. We found a tentative rate of 0.05 novae yr$^{-1}$ GC$^{-1}$ that is about an order of magnitude higher than expected from stellar evolution or upper limits from earlier optical surveys. Complementing analyses of the recent WeCAPP optical survey provided a lower nova rate (by about an order of magnitude), but did not contain one of the two known optical GC novae in its field of view and is still consistent with the X-ray estimate on the 95\% level. Furthermore, we estimated that the observed luminosity-specific nova rate is at least (on the 95\% confidence level) a factor of three higher in \m31 GCs than for the (similarly old) bulge population. These results further underline the need for additional processes in GCs, leading to dynamical binary formation or more effective accretion onto compact objects.

Unlike \novak, most novae with an SSS counterpart in \m31 were discovered in the optical before their X-ray emission was detected \citep[see e.g.][]{2005A&A...442..879P,2007A&A...465..375P,2010A&A...523A..89H,2011A&A...533A..52H}. This is due to quiescent \m31 novae being too faint to be observed with small optical telescopes and nova surveys therefore search for the eponymous new stars. For novae in relatively bright GCs, though, the outburst signature is much more subtle (see Fig.\,\ref{fig:lc_nova}) and can easily go unnoticed, as happened for \nova until its discovery as an SSS. X-ray monitoring surveys with high cadence provide a powerful method to discover CNe in GCs and study their individual properties and overall outburst rates.

\begin{acknowledgements}
The anonymous referee is acknowledged for constructive comments that helped to improve the clarity of the paper. We would like to thank K. Lutz for her help in setting up the optical data reduction pipeline. The \xmm project is supported by the Bundesministerium f\"{u}r Wirtschaft und Technologie / Deutsches Zentrum f\"{u}r Luft- und Raumfahrt (BMWI/DLR FKZ 50 OX 0001) and the Max Planck Society. We would like to thank the \swift team for the scheduling of the ToO observations. M. Henze acknowledges support from the BMWI/DLR, FKZ 50 OR 1010. The work of K. Hornoch was supported by the project RVO:67985815. We thank M. Wolf, P. Hornochov\'a, P. Ku\v{s}nir\'ak, and P. Zasche for their assistance with acquiring of the observations at the Ond\v{r}ejov observatory. The PS1 Surveys have been made possible through contributions of the Institute for Astronomy at the University of Hawaii in Manoa, the Pan-STARRS Project Office, the Max Planck Society and its participating institutes, the Max Planck Institute for Astronomy, Heidelberg and the Max Planck Institute for Extraterrestrial Physics, Garching, the Johns Hopkins University, the University of Durham, the University of Edinburgh, the Queens University of Belfast, the Harvard-Smithsonian Center for Astrophysics, and the Los Cumbres Observatory Global Telescope Network, Incorporated, and the National Central University of Taiwan.

\end{acknowledgements}

\bibliographystyle{aa}

%
\begin{table*}[ht]
\caption{Optical observations of \bol with small telescopes.}
\label{tab:obs_opt}
\begin{center}
\begin{tabular}{lccccc}\hline\hline \noalign{\smallskip}
Observatory $^a$ & Date & MJD & $\Delta t ^b$ &  Mag. GC $^c$ & Mag. nova $^d$\\
& [UT] & [d] & [d] & [mag] & [mag]\\ \hline \noalign{\smallskip}
Ond\v{r}ejov & 2010-10-06T22:22:05 & 55475.93 & -4.58 & $16.72 \pm 0.02$ & - \\
Super-LOTIS & 2010-10-07T03:02:50 & 55476.13 & -4.38 & $16.67 \pm 0.48$ & - \\
Ond\v{r}ejov & 2010-10-08T19:37:55 & 55477.82 & -2.69 & $16.69 \pm 0.02$ & - \\
Super-LOTIS & 2010-10-09T03:00:43 & 55478.13 & -2.38 & $16.53 \pm 0.16$ & - \\
Miyaki-Argenteus & 2010-10-09T14:19:17 & 55478.60 & -1.91 & $16.62 \pm 0.16$ & - \\
Miyaki-Argenteus & 2010-10-10T15:01:01 & 55479.63 & -0.88 & $16.69 \pm 0.17$ & - \\
Ond\v{r}ejov & 2010-10-10T19:20:38 & 55479.81 & -0.70 & $16.70 \pm 0.02$ & - \\
Super-LOTIS & 2010-10-11T02:12:57 & 55480.09 & -0.42 & $16.65 \pm 0.13$ & - \\
Xingming & 2010-10-11T13:21:30 & 55480.56 & 0.05 & $16.31 \pm 0.22$ & $17.60 \pm 0.96$ \\
Super-LOTIS & 2010-10-12T11:15:42 & 55481.47 & 0.96 & $15.62 \pm 0.14$ & $16.12 \pm 0.28$ \\
Xingming & 2010-10-12T15:02:56 & 55481.63 & 1.12 & $15.67 \pm 0.26$ & $16.21 \pm 0.48$ \\
Ond\v{r}ejov & 2010-10-12T19:30:43 & 55481.81 & 1.30 & $15.90 \pm 0.02$ & $16.61 \pm 0.13$ \\
Xingming & 2010-10-13T13:24:07 & 55482.56 & 2.05 & $16.18 \pm 0.23$ & $17.23 \pm 0.77$ \\
Ond\v{r}ejov & 2010-10-13T20:39:50 & 55482.86 & 2.35 & $16.19 \pm 0.02$ & $17.26 \pm 0.22$ \\
Xingming & 2010-10-14T14:01:17 & 55483.58 & 3.07 & $16.19 \pm 0.29$ & $17.25 \pm 0.94$ \\
Miyaki-Argenteus & 2010-10-15T13:35:27 & 55484.57 & 4.06 & $16.30 \pm 0.19$ & $17.57 \pm 0.83$ \\
Itagaki & 2010-10-16T09:22:11 & 55485.39 & 4.88 & $16.42 \pm 0.29$ & $18.04 \pm 1.62$ \\
Xingming & 2010-10-16T13:35:54 & 55485.57 & 5.06 & $16.42 \pm 0.21$ & $18.03 \pm 1.26$ \\
Miyaki-Argenteus & 2010-10-16T14:59:20 & 55485.62 & 5.11 & $16.48 \pm 0.15$ & $18.32 \pm 1.25$ \\
Miyaki-Argenteus & 2010-10-17T12:57:21 & 55486.54 & 6.03 & $16.57 \pm 0.14$ & $18.94 \pm 2.04$ \\
Xingming & 2010-10-17T15:32:06 & 55486.65 & 6.14 & $16.45 \pm 0.25$ & $18.19 \pm 1.65$ \\
Itagaki & 2010-10-17T15:43:06 & 55486.65 & 6.14 & $16.49 \pm 0.21$ & $18.37 \pm 1.66$ \\
Miyaki-Argenteus & 2010-10-18T12:05:45 & 55487.50 & 6.99 & $16.70 \pm 0.18$ & - \\
Ond\v{r}ejov & 2010-10-19T00:14:24 & 55488.01 & 7.50 & $16.54 \pm 0.02$ & $18.70 \pm 0.78$ \\
Itagaki & 2010-10-22T08:47:15 & 55491.37 & 10.86 & $16.58 \pm 0.29$ & - \\
Super-LOTIS & 2010-10-25T05:14:32 & 55494.22 & 13.71 & $16.47 \pm 0.14$ & - \\
Miyaki-Argenteus & 2010-10-26T13:11:55 & 55495.55 & 15.04 & $16.48 \pm 0.15$ & - \\
Super-LOTIS & 2010-10-27T03:45:12 & 55496.16 & 15.65 & $16.67 \pm 0.14$ & - \\
Miyaki-Argenteus & 2010-10-27T15:24:36 & 55496.64 & 16.13 & $16.61 \pm 0.17$ & - \\
Super-LOTIS & 2010-10-28T02:39:48 & 55497.11 & 16.60 & $16.70 \pm 0.14$ & - \\
Ond\v{r}ejov & 2010-10-28T16:52:19 & 55497.70 & 17.19 & $16.68 \pm 0.02$ & - \\
Super-LOTIS & 2010-10-29T01:55:36 & 55498.08 & 17.57 & $16.64 \pm 0.13$ & - \\
Miyaki-Argenteus & 2010-10-29T11:28:38 & 55498.48 & 17.97 & $16.93 \pm 0.19$ & - \\
Ond\v{r}ejov & 2010-10-29T17:03:50 & 55498.71 & 18.20 & $16.70 \pm 0.02$ & - \\
Super-LOTIS & 2010-10-30T02:04:25 & 55499.09 & 18.58 & $16.67 \pm 0.12$ & - \\
Super-LOTIS & 2010-11-01T05:35:20 & 55501.23 & 20.72 & $16.65 \pm 0.14$ & - \\
Miyaki-Argenteus & 2010-11-01T12:35:39 & 55501.52 & 21.01 & $16.79 \pm 0.16$ & - \\
Super-LOTIS & 2010-11-02T02:35:30 & 55502.11 & 21.60 & $16.62 \pm 0.13$ & - \\
Miyaki-Argenteus & 2010-11-02T13:43:47 & 55502.57 & 22.06 & $16.53 \pm 0.14$ & - \\
Super-LOTIS & 2010-11-03T02:34:53 & 55503.11 & 22.60 & $16.55 \pm 0.13$ & - \\
Miyaki-Argenteus & 2010-11-03T12:01:05 & 55503.50 & 22.99 & $16.50 \pm 0.15$ & - \\
Super-LOTIS & 2010-11-04T01:51:30 & 55504.08 & 23.57 & $16.57 \pm 0.12$ & - \\
Miyaki-Argenteus & 2010-11-04T12:59:54 & 55504.54 & 24.03 & $16.68 \pm 0.13$ & - \\
Super-LOTIS & 2010-11-05T01:50:35 & 55505.08 & 24.57 & $16.63 \pm 0.13$ & - \\
Miyaki-Argenteus & 2010-11-05T11:26:58 & 55505.48 & 24.97 & $16.68 \pm 0.14$ & - \\
Super-LOTIS & 2010-11-09T02:31:25 & 55509.11 & 28.60 & $16.64 \pm 0.13$ & - \\
\hline
\end{tabular}
\end{center}
\noindent
Notes:\hspace{0.1cm} $^a $: See Sect.\,\ref{sec:obs_opt} for a description of the telescopes used; $^b $: Time in days after the optical nova outburst on MJD = 55480.32; $^c $: $R$ band magnitude of \bolk, see also Fig.\,\ref{fig:lc_opt}; $^d $: $R$ band magnitude of \nova assuming a quiescence magnitude for \bol of $R = 16.7$ mag, see also Fig.\,\ref{fig:lc_nova}.\\
\end{table*}

\clearpage

%
\begin{table}[ht]
\caption{Optical PAndromeda observations of \bolk.}
\label{tab:obs_pan2}
\begin{center}
\begin{tabular}{lcccc}\hline\hline \noalign{\smallskip}
Date & MJD & $\Delta t ^a$ &  Mag. GC $^b$ & Filter\\
 {[}UT] & [d] & [d] & [mag] & \\ \hline \noalign{\smallskip}
2010-07-24.46 & 55401.46 & -79.05 & $16.899 \pm 0.023$ & r \\
2010-07-24.59 & 55401.59 & -78.92 & $16.936 \pm 0.030$ & r \\
2010-07-25.47 & 55402.47 & -78.04 & $16.903 \pm 0.022$ & r \\
2010-07-25.60 & 55402.60 & -77.91 & $16.892 \pm 0.019$ & r \\
2010-07-27.46 & 55404.46 & -76.05 & $16.894 \pm 0.031$ & r \\
2010-07-27.58 & 55404.58 & -75.93 & $16.901 \pm 0.014$ & r \\
2010-07-28.47 & 55405.47 & -75.04 & $16.885 \pm 0.012$ & r \\
2010-07-28.59 & 55405.59 & -74.92 & $16.894 \pm 0.016$ & r \\
2010-07-29.48 & 55406.48 & -74.03 & $16.909 \pm 0.018$ & r \\
2010-07-29.60 & 55406.60 & -73.91 & $16.900 \pm 0.016$ & r \\
2010-07-30.49 & 55407.49 & -73.02 & $16.888 \pm 0.020$ & r \\
2010-07-30.62 & 55407.62 & -72.89 & $16.865 \pm 0.013$ & r \\
2010-07-31.48 & 55408.48 & -72.03 & $16.881 \pm 0.032$ & r \\
2010-07-31.60 & 55408.60 & -71.91 & $16.880 \pm 0.015$ & r \\
2010-08-01.48 & 55409.48 & -71.03 & $16.892 \pm 0.016$ & r \\
2010-08-01.60 & 55409.60 & -70.91 & $16.878 \pm 0.018$ & r \\
2010-08-02.50 & 55410.50 & -70.01 & $16.879 \pm 0.018$ & r \\
2010-08-02.62 & 55410.62 & -69.89 & $16.879 \pm 0.018$ & r \\
2010-08-03.48 & 55411.48 & -69.03 & $16.901 \pm 0.019$ & r \\
2010-08-03.61 & 55411.61 & -68.90 & $16.894 \pm 0.020$ & r \\
2010-08-04.49 & 55412.49 & -68.02 & $16.894 \pm 0.014$ & r \\
2010-08-04.61 & 55412.61 & -67.90 & $16.881 \pm 0.015$ & r \\
2010-08-05.48 & 55413.48 & -67.03 & $16.898 \pm 0.013$ & r \\
2010-08-14.47 & 55422.47 & -58.04 & $16.903 \pm 0.018$ & r \\
2010-08-14.62 & 55422.62 & -57.89 & $16.901 \pm 0.020$ & r \\
2010-08-15.49 & 55423.49 & -57.02 & $16.895 \pm 0.018$ & r \\
2010-08-15.62 & 55423.62 & -56.89 & $16.886 \pm 0.019$ & r \\
2010-08-16.49 & 55424.49 & -56.02 & $16.892 \pm 0.012$ & r \\
2010-08-16.62 & 55424.62 & -55.89 & $16.878 \pm 0.017$ & r \\
2010-08-17.49 & 55425.49 & -55.02 & $16.905 \pm 0.012$ & r \\
2010-08-17.61 & 55425.61 & -54.90 & $16.881 \pm 0.013$ & r \\
2010-08-18.48 & 55426.48 & -54.03 & $16.900 \pm 0.013$ & r \\
2010-08-18.61 & 55426.61 & -53.90 & $16.901 \pm 0.018$ & r \\
2010-08-19.47 & 55427.47 & -53.04 & $16.897 \pm 0.018$ & r \\
2010-08-19.62 & 55427.62 & -52.89 & $16.871 \pm 0.013$ & r \\
2010-08-20.47 & 55428.47 & -52.04 & $16.893 \pm 0.019$ & r \\
2010-08-20.60 & 55428.60 & -51.91 & $16.904 \pm 0.015$ & r \\
2010-08-21.48 & 55429.48 & -51.03 & $16.897 \pm 0.015$ & r \\
2010-08-21.61 & 55429.61 & -50.90 & $16.890 \pm 0.014$ & r \\
2010-08-22.48 & 55430.48 & -50.03 & $16.890 \pm 0.011$ & r \\
2010-08-22.59 & 55430.59 & -49.92 & $16.888 \pm 0.012$ & r \\
2010-08-23.49 & 55431.49 & -49.02 & $16.905 \pm 0.020$ & r \\
2010-08-24.47 & 55432.47 & -48.04 & $16.886 \pm 0.012$ & r \\
2010-08-24.58 & 55432.58 & -47.93 & $16.880 \pm 0.016$ & r \\
2010-08-25.45 & 55433.45 & -47.06 & $16.906 \pm 0.017$ & r \\
2010-08-25.59 & 55433.59 & -46.92 & $16.878 \pm 0.023$ & r \\
2010-08-26.48 & 55434.48 & -46.03 & $16.885 \pm 0.015$ & r \\
2010-08-26.63 & 55434.63 & -45.88 & $16.905 \pm 0.016$ & r \\
2010-08-27.48 & 55435.48 & -45.03 & $16.895 \pm 0.023$ & r \\
2010-08-27.60 & 55435.60 & -44.91 & $16.895 \pm 0.016$ & r \\
2010-08-28.48 & 55436.48 & -44.03 & $16.887 \pm 0.016$ & r \\
2010-08-28.60 & 55436.60 & -43.91 & $16.877 \pm 0.018$ & r \\
2010-08-31.46 & 55439.46 & -41.05 & $16.879 \pm 0.014$ & r \\
2010-08-31.58 & 55439.58 & -40.93 & $16.862 \pm 0.015$ & r \\
2010-09-01.48 & 55440.48 & -40.03 & $16.861 \pm 0.012$ & r \\
2010-09-01.62 & 55440.62 & -39.89 & $16.882 \pm 0.013$ & r \\
2010-09-02.49 & 55441.49 & -39.02 & $16.863 \pm 0.013$ & r \\
2010-09-02.61 & 55441.61 & -38.90 & $16.878 \pm 0.011$ & r \\
2010-09-03.48 & 55442.48 & -38.03 & $16.857 \pm 0.012$ & r \\
2010-09-03.59 & 55442.59 & -37.92 & $16.904 \pm 0.015$ & r \\
\hline
\end{tabular}
\end{center}
\noindent
Notes:\hspace{0.1cm} $^a $: Time in days after the optical nova outburst on MJD = 55480.51; $^b $: $r$ or $i$ band magnitude of \bolk, see also Fig.\,\ref{fig:lc_pan}.\
\end{table}
%

%
\begin{table}[ht]
\addtocounter{table}{-1}
\caption{continued.}
\begin{center}
\begin{tabular}{lcccc}\hline\hline \noalign{\smallskip}
Date & MJD & $\Delta t ^a$ &  Mag. GC $^b$ & Filter\\
 {[}UT] & [d] & [d] & [mag] & \\ \hline \noalign{\smallskip}
2010-09-04.45 & 55443.45 & -37.06 & $16.865 \pm 0.012$ & r \\
2010-09-05.45 & 55444.45 & -36.06 & $16.884 \pm 0.012$ & r \\
2010-09-05.57 & 55444.57 & -35.94 & $16.886 \pm 0.014$ & r \\
2010-09-06.45 & 55445.45 & -35.06 & $16.886 \pm 0.017$ & r \\
2010-09-06.59 & 55445.59 & -34.92 & $16.901 \pm 0.014$ & r \\
2010-09-07.44 & 55446.44 & -34.07 & $16.890 \pm 0.014$ & r \\
2010-09-07.58 & 55446.58 & -33.93 & $16.886 \pm 0.013$ & r \\
2010-09-08.44 & 55447.44 & -33.07 & $16.880 \pm 0.011$ & r \\
2010-09-08.57 & 55447.57 & -32.94 & $16.879 \pm 0.014$ & r \\
2010-09-09.42 & 55448.42 & -32.09 & $16.894 \pm 0.016$ & r \\
2010-09-09.54 & 55448.54 & -31.97 & $16.892 \pm 0.019$ & r \\
2010-09-10.43 & 55449.43 & -31.08 & $16.889 \pm 0.021$ & r \\
2010-09-10.56 & 55449.56 & -30.95 & $16.909 \pm 0.019$ & r \\
2010-09-11.42 & 55450.42 & -30.09 & $16.897 \pm 0.024$ & r \\
2010-09-12.41 & 55451.41 & -29.10 & $16.902 \pm 0.015$ & r \\
2010-09-12.55 & 55451.55 & -28.96 & $16.888 \pm 0.015$ & r \\
2010-09-13.42 & 55452.42 & -28.09 & $16.881 \pm 0.014$ & r \\
2010-09-13.56 & 55452.56 & -27.95 & $16.897 \pm 0.018$ & r \\
2010-09-14.42 & 55453.42 & -27.09 & $16.894 \pm 0.012$ & r \\
2010-09-14.53 & 55453.53 & -26.98 & $16.906 \pm 0.017$ & r \\
2010-09-15.41 & 55454.41 & -26.10 & $16.895 \pm 0.015$ & r \\
2010-09-15.53 & 55454.53 & -25.98 & $16.894 \pm 0.018$ & r \\
2010-09-16.41 & 55455.41 & -25.10 & $16.903 \pm 0.013$ & r \\
2010-09-16.54 & 55455.54 & -24.97 & $16.898 \pm 0.014$ & r \\
2010-09-17.41 & 55456.41 & -24.10 & $16.901 \pm 0.013$ & r \\
2010-09-17.55 & 55456.55 & -23.96 & $16.871 \pm 0.014$ & r \\
2010-09-18.39 & 55457.39 & -23.12 & $16.894 \pm 0.013$ & r \\
2010-09-18.53 & 55457.53 & -22.98 & $16.899 \pm 0.019$ & r \\
2010-09-24.38 & 55463.38 & -17.13 & $16.891 \pm 0.020$ & r \\
2010-09-24.52 & 55463.52 & -16.99 & $16.894 \pm 0.016$ & r \\
2010-09-27.37 & 55466.37 & -14.14 & $16.911 \pm 0.020$ & r \\
2010-09-27.50 & 55466.50 & -14.01 & $16.903 \pm 0.023$ & r \\
2010-09-28.36 & 55467.36 & -13.15 & $16.871 \pm 0.025$ & r \\
2010-09-28.51 & 55467.51 & -13.00 & $16.890 \pm 0.020$ & r \\
2010-09-29.37 & 55468.37 & -12.14 & $16.898 \pm 0.021$ & r \\
2010-09-29.54 & 55468.54 & -11.97 & $16.957 \pm 0.037$ & r \\
2010-10-03.37 & 55472.37 & -8.14 & $16.897 \pm 0.020$ & r \\
2010-10-03.49 & 55472.49 & -8.02 & $16.904 \pm 0.014$ & r \\
2010-10-06.35 & 55475.35 & -5.16 & $16.919 \pm 0.019$ & r \\
2010-10-06.49 & 55475.49 & -5.02 & $16.903 \pm 0.017$ & r \\
2010-10-07.35 & 55476.35 & -4.16 & $16.889 \pm 0.012$ & r \\
2010-10-07.48 & 55476.48 & -4.03 & $16.894 \pm 0.012$ & r \\
2010-10-08.36 & 55477.36 & -3.15 & $16.885 \pm 0.012$ & r \\
2010-10-08.49 & 55477.49 & -3.02 & $16.892 \pm 0.013$ & r \\
2010-10-09.35 & 55478.35 & -2.16 & $16.896 \pm 0.011$ & r \\
2010-10-09.49 & 55478.49 & -2.02 & $16.906 \pm 0.014$ & r \\
2010-10-10.34 & 55479.34 & -1.17 & $16.905 \pm 0.017$ & r \\
2010-10-10.48 & 55479.48 & -1.03 & $16.928 \pm 0.017$ & r \\
2010-10-11.34 & 55480.34 & -0.17 & $16.903 \pm 0.020$ & r \\
2010-10-11.46 & 55480.46 & -0.05 & $16.880 \pm 0.016$ & r \\
2010-10-12.34 & 55481.34 & 0.83 & $15.512 \pm 0.009$ & r \\
2010-10-12.46 & 55481.46 & 0.95 & $15.839 \pm 0.014$ & r \\
2010-10-13.51 & 55482.51 & 2.00 & $16.220 \pm 0.009$ & r \\
2010-10-14.33 & 55483.33 & 2.82 & $16.296 \pm 0.010$ & r \\
2010-10-14.47 & 55483.47 & 2.96 & $16.454 \pm 0.012$ & r \\
2010-10-15.34 & 55484.34 & 3.83 & $16.442 \pm 0.011$ & r \\
2010-10-15.46 & 55484.46 & 3.95 & $16.557 \pm 0.010$ & r \\
2010-10-18.32 & 55487.32 & 6.81 & $16.714 \pm 0.019$ & r \\
2010-10-18.45 & 55487.45 & 6.94 & $16.730 \pm 0.012$ & r \\
2010-10-20.32 & 55489.32 & 8.81 & $16.754 \pm 0.020$ & r \\
2010-10-27.31 & 55496.31 & 15.80 & $16.873 \pm 0.051$ & r \\
2010-10-29.31 & 55498.31 & 17.80 & $16.861 \pm 0.012$ & r \\
2010-10-29.43 & 55498.43 & 17.92 & $16.896 \pm 0.016$ & r \\
\hline
\end{tabular}
\end{center}
\end{table}
%

%
\begin{table}[ht]
\addtocounter{table}{-1}
\caption{continued.}
\begin{center}
\begin{tabular}{lcccc}\hline\hline \noalign{\smallskip}
Date & MJD & $\Delta t ^a$ &  Mag. GC $^b$ & Filter\\
 {[}UT] & [d] & [d] & [mag] & \\ \hline \noalign{\smallskip}
2010-10-30.31 & 55499.31 & 18.80 & $16.877 \pm 0.012$ & r \\
2010-10-30.43 & 55499.43 & 18.92 & $16.873 \pm 0.014$ & r \\
2010-10-31.31 & 55500.31 & 19.80 & $16.866 \pm 0.011$ & r \\
2010-10-31.44 & 55500.44 & 19.93 & $16.872 \pm 0.011$ & r \\
2010-11-01.27 & 55501.27 & 20.76 & $16.879 \pm 0.013$ & r \\
2010-11-01.43 & 55501.43 & 20.92 & $16.893 \pm 0.012$ & r \\
2010-11-02.28 & 55502.27 & 21.76 & $16.881 \pm 0.016$ & r \\
2010-11-02.41 & 55502.41 & 21.90 & $16.882 \pm 0.010$ & r \\
2010-11-03.27 & 55503.27 & 22.76 & $16.897 \pm 0.014$ & r \\
2010-11-03.40 & 55503.40 & 22.89 & $16.880 \pm 0.012$ & r \\
2010-11-07.29 & 55507.29 & 26.78 & $16.914 \pm 0.021$ & r \\
2010-11-07.41 & 55507.41 & 26.90 & $16.889 \pm 0.011$ & r \\
2010-11-08.28 & 55508.28 & 27.77 & $16.876 \pm 0.018$ & r \\
2010-11-08.40 & 55508.40 & 27.89 & $16.885 \pm 0.013$ & r \\
2010-11-10.27 & 55510.27 & 29.76 & $16.885 \pm 0.013$ & r \\
2010-11-10.41 & 55510.41 & 29.90 & $16.896 \pm 0.014$ & r \\
2010-11-11.28 & 55511.28 & 30.77 & $16.884 \pm 0.014$ & r \\
2010-11-11.40 & 55511.40 & 30.89 & $16.892 \pm 0.011$ & r \\
2010-11-12.25 & 55512.25 & 31.74 & $16.890 \pm 0.027$ & r \\
2010-11-14.26 & 55514.26 & 33.75 & $16.917 \pm 0.018$ & r \\
2010-11-16.24 & 55516.24 & 35.73 & $16.910 \pm 0.014$ & r \\
2010-11-16.38 & 55516.38 & 35.86 & $16.893 \pm 0.011$ & r \\
2010-11-18.25 & 55518.25 & 37.74 & $16.896 \pm 0.013$ & r \\
2010-11-18.39 & 55518.39 & 37.88 & $16.877 \pm 0.012$ & r \\
2010-11-21.22 & 55521.22 & 40.71 & $16.897 \pm 0.016$ & r \\
2010-11-21.39 & 55521.39 & 40.88 & $16.916 \pm 0.016$ & r \\
2010-11-23.24 & 55523.24 & 42.73 & $16.920 \pm 0.013$ & r \\
2010-11-23.37 & 55523.37 & 42.86 & $16.905 \pm 0.017$ & r \\
2010-11-30.21 & 55530.21 & 49.70 & $16.887 \pm 0.030$ & r \\
2010-11-30.36 & 55530.36 & 49.85 & $16.903 \pm 0.028$ & r \\
2010-12-03.23 & 55533.23 & 52.72 & $16.913 \pm 0.174$ & r \\
2010-12-04.23 & 55534.23 & 53.72 & $16.895 \pm 0.010$ & r \\
2010-12-04.36 & 55534.36 & 53.85 & $16.894 \pm 0.012$ & r \\
2010-12-05.23 & 55535.23 & 54.72 & $16.909 \pm 0.017$ & r \\
2010-12-05.35 & 55535.35 & 54.84 & $16.873 \pm 0.021$ & r \\
2010-12-25.23 & 55555.23 & 74.72 & $16.882 \pm 0.020$ & r \\
2010-12-25.33 & 55555.33 & 74.82 & $16.893 \pm 0.023$ & r \\
2010-12-27.22 & 55557.22 & 76.71 & $16.883 \pm 0.020$ & r \\
2010-12-27.33 & 55557.33 & 76.82 & $16.907 \pm 0.018$ & r \\
2011-07-25.61 & 55767.61 & 287.10 & $16.960 \pm 0.015$ & r \\
2011-07-26.61 & 55768.61 & 288.10 & $16.951 \pm 0.014$ & r \\
2011-07-27.60 & 55769.60 & 289.09 & $16.957 \pm 0.007$ & r \\
2011-07-28.59 & 55770.59 & 290.08 & $16.952 \pm 0.013$ & r \\
2011-07-31.53 & 55773.53 & 293.02 & $16.951 \pm 0.010$ & r \\
2011-08-01.55 & 55774.55 & 294.04 & $16.946 \pm 0.008$ & r \\
2011-08-02.56 & 55775.56 & 295.05 & $16.948 \pm 0.008$ & r \\
2011-08-03.61 & 55776.61 & 296.10 & $16.963 \pm 0.008$ & r \\
2011-08-04.47 & 55777.47 & 296.96 & $16.911 \pm 0.012$ & r \\
2011-08-04.61 & 55777.61 & 297.10 & $16.961 \pm 0.012$ & r \\
2011-08-06.48 & 55779.48 & 298.97 & $16.942 \pm 0.010$ & r \\
2011-08-06.61 & 55779.61 & 299.10 & $16.959 \pm 0.008$ & r \\
2011-08-09.48 & 55782.48 & 301.97 & $16.916 \pm 0.024$ & r \\
2011-08-09.61 & 55782.61 & 302.10 & $16.939 \pm 0.014$ & r \\
2011-08-10.47 & 55783.47 & 302.96 & $16.943 \pm 0.010$ & r \\
2011-08-10.62 & 55783.62 & 303.11 & $16.913 \pm 0.008$ & r \\
2011-08-11.48 & 55784.48 & 303.97 & $16.947 \pm 0.011$ & r \\
2011-08-11.59 & 55784.59 & 304.08 & $16.895 \pm 0.009$ & r \\
2011-08-12.48 & 55785.48 & 304.97 & $16.958 \pm 0.009$ & r \\
2010-07-23.53 & 55400.53 & -79.98 & $16.713 \pm 0.016$ & i \\
2010-07-24.45 & 55401.45 & -79.06 & $16.702 \pm 0.022$ & i \\
2010-07-24.58 & 55401.58 & -78.93 & $16.690 \pm 0.027$ & i \\
2010-07-25.47 & 55402.47 & -78.04 & $16.702 \pm 0.021$ & i \\
2010-07-25.60 & 55402.60 & -77.91 & $16.684 \pm 0.016$ & i \\
\hline
\end{tabular}
\end{center}
\end{table}
%

%
\begin{table}[ht]
\addtocounter{table}{-1}
\caption{continued.}
\begin{center}
\begin{tabular}{lcccc}\hline\hline \noalign{\smallskip}
Date & MJD & $\Delta t ^a$ &  Mag. GC $^b$ & Filter\\
 {[}UT] & [d] & [d] & [mag] & \\ \hline \noalign{\smallskip}
2010-07-27.46 & 55404.46 & -76.05 & $16.694 \pm 0.038$ & i \\
2010-07-27.58 & 55404.58 & -75.93 & $16.695 \pm 0.011$ & i \\
2010-07-28.46 & 55405.46 & -75.05 & $16.693 \pm 0.011$ & i \\
2010-07-28.58 & 55405.58 & -74.93 & $16.692 \pm 0.011$ & i \\
2010-07-29.47 & 55406.47 & -74.04 & $16.704 \pm 0.019$ & i \\
2010-07-29.59 & 55406.59 & -73.92 & $16.677 \pm 0.012$ & i \\
2010-07-30.49 & 55407.49 & -73.02 & $16.693 \pm 0.013$ & i \\
2010-07-30.61 & 55407.61 & -72.90 & $16.669 \pm 0.012$ & i \\
2010-07-31.47 & 55408.47 & -72.04 & $16.702 \pm 0.029$ & i \\
2010-07-31.60 & 55408.60 & -71.91 & $16.697 \pm 0.012$ & i \\
2010-08-01.48 & 55409.48 & -71.03 & $16.685 \pm 0.013$ & i \\
2010-08-01.59 & 55409.59 & -70.92 & $16.672 \pm 0.013$ & i \\
2010-08-02.50 & 55410.50 & -70.01 & $16.693 \pm 0.017$ & i \\
2010-08-02.62 & 55410.62 & -69.89 & $16.722 \pm 0.024$ & i \\
2010-08-03.48 & 55411.48 & -69.03 & $16.710 \pm 0.014$ & i \\
2010-08-03.60 & 55411.60 & -68.91 & $16.676 \pm 0.014$ & i \\
2010-08-04.48 & 55412.48 & -68.03 & $16.700 \pm 0.012$ & i \\
2010-08-04.60 & 55412.60 & -67.91 & $16.678 \pm 0.011$ & i \\
2010-08-05.47 & 55413.47 & -67.04 & $16.695 \pm 0.011$ & i \\
2010-08-05.61 & 55413.61 & -66.90 & $16.692 \pm 0.014$ & i \\
2010-08-22.47 & 55430.47 & -50.04 & $16.696 \pm 0.009$ & i \\
2010-08-22.58 & 55430.58 & -49.93 & $16.681 \pm 0.012$ & i \\
2010-08-23.49 & 55431.49 & -49.02 & $16.703 \pm 0.012$ & i \\
2010-08-24.46 & 55432.46 & -48.05 & $16.691 \pm 0.010$ & i \\
2010-08-24.57 & 55432.57 & -47.94 & $16.680 \pm 0.019$ & i \\
2010-08-25.45 & 55433.45 & -47.06 & $16.718 \pm 0.016$ & i \\
2010-08-25.58 & 55433.58 & -46.93 & $16.685 \pm 0.011$ & i \\
2010-08-26.48 & 55434.48 & -46.03 & $16.692 \pm 0.012$ & i \\
2010-08-26.61 & 55434.61 & -45.90 & $16.670 \pm 0.018$ & i \\
2010-08-27.47 & 55435.47 & -45.04 & $16.689 \pm 0.015$ & i \\
2010-08-27.59 & 55435.59 & -44.92 & $16.684 \pm 0.013$ & i \\
2010-08-28.47 & 55436.47 & -44.04 & $16.684 \pm 0.015$ & i \\
2010-08-28.60 & 55436.60 & -43.91 & $16.671 \pm 0.011$ & i \\
2010-08-31.46 & 55439.46 & -41.05 & $16.685 \pm 0.012$ & i \\
2010-08-31.59 & 55439.59 & -40.92 & $16.683 \pm 0.011$ & i \\
2010-09-01.48 & 55440.48 & -40.03 & $16.688 \pm 0.011$ & i \\
2010-09-01.61 & 55440.61 & -39.90 & $16.692 \pm 0.011$ & i \\
2010-09-02.48 & 55441.48 & -39.03 & $16.668 \pm 0.012$ & i \\
2010-09-02.60 & 55441.60 & -38.91 & $16.662 \pm 0.010$ & i \\
2010-09-03.47 & 55442.47 & -38.04 & $16.676 \pm 0.010$ & i \\
2010-09-03.59 & 55442.59 & -37.92 & $16.675 \pm 0.012$ & i \\
2010-09-04.45 & 55443.45 & -37.06 & $16.690 \pm 0.010$ & i \\
2010-09-05.45 & 55444.45 & -36.06 & $16.690 \pm 0.011$ & i \\
2010-09-05.57 & 55444.57 & -35.94 & $16.684 \pm 0.011$ & i \\
2010-09-06.45 & 55445.45 & -35.06 & $16.682 \pm 0.012$ & i \\
2010-09-06.58 & 55445.58 & -34.93 & $16.689 \pm 0.012$ & i \\
2010-09-07.44 & 55446.44 & -34.07 & $16.689 \pm 0.011$ & i \\
2010-09-07.58 & 55446.58 & -33.93 & $16.684 \pm 0.012$ & i \\
2010-09-28.37 & 55467.37 & -13.14 & $16.693 \pm 0.012$ & i \\
2010-09-28.51 & 55467.51 & -13.00 & $16.659 \pm 0.026$ & i \\
2010-09-29.37 & 55468.37 & -12.14 & $16.685 \pm 0.024$ & i \\
2010-09-29.55 & 55468.55 & -11.96 & $16.685 \pm 0.027$ & i \\
2010-10-03.37 & 55472.37 & -8.14 & $16.700 \pm 0.019$ & i \\
2010-10-03.50 & 55472.50 & -8.01 & $16.694 \pm 0.012$ & i \\
2010-10-06.35 & 55475.35 & -5.16 & $16.706 \pm 0.016$ & i \\
2010-10-06.48 & 55475.48 & -5.03 & $16.703 \pm 0.013$ & i \\
2010-10-09.35 & 55478.35 & -2.16 & $16.692 \pm 0.011$ & i \\
2010-10-09.50 & 55478.50 & -2.01 & $16.690 \pm 0.011$ & i \\
2010-10-10.35 & 55479.35 & -1.16 & $16.706 \pm 0.015$ & i \\
2010-10-10.49 & 55479.49 & -1.02 & $16.692 \pm 0.016$ & i \\
2010-10-11.35 & 55480.35 & -0.16 & $16.699 \pm 0.018$ & i \\
2010-10-11.46 & 55480.46 & -0.05 & $16.684 \pm 0.015$ & i \\
2010-10-12.35 & 55481.35 & 0.84 & $15.602 \pm 0.009$ & i \\
\hline
\end{tabular}
\end{center}
\end{table}
%

%
\begin{table}[ht]
\addtocounter{table}{-1}
\caption{continued.}
\begin{center}
\begin{tabular}{lcccc}\hline\hline \noalign{\smallskip}
Date & MJD & $\Delta t ^a$ &  Mag. GC $^b$ & Filter\\
 {[}UT] & [d] & [d] & [mag] & \\ \hline \noalign{\smallskip}
2010-10-12.46 & 55481.46 & 0.95 & $15.702 \pm 0.010$ & i \\
2010-10-13.33 & 55482.33 & 1.82 & $16.118 \pm 0.032$ & i \\
2010-10-13.50 & 55482.50 & 1.99 & $16.169 \pm 0.009$ & i \\
2010-10-14.32 & 55483.32 & 2.81 & $16.349 \pm 0.011$ & i \\
2010-10-14.46 & 55483.46 & 2.95 & $16.415 \pm 0.010$ & i \\
2010-10-15.33 & 55484.33 & 3.82 & $16.491 \pm 0.010$ & i \\
2010-10-15.46 & 55484.46 & 3.95 & $16.525 \pm 0.010$ & i \\
2010-10-18.33 & 55487.33 & 6.82 & $16.651 \pm 0.020$ & i \\
2010-10-18.45 & 55487.45 & 6.94 & $16.635 \pm 0.011$ & i \\
2010-10-20.31 & 55489.31 & 8.80 & $16.660 \pm 0.024$ & i \\
2010-10-27.30 & 55496.30 & 15.79 & $16.610 \pm 0.055$ & i \\
2010-10-29.30 & 55498.30 & 17.79 & $16.688 \pm 0.010$ & i \\
2010-10-29.43 & 55498.43 & 17.92 & $16.667 \pm 0.016$ & i \\
2010-10-30.30 & 55499.30 & 18.79 & $16.693 \pm 0.013$ & i \\
2010-10-30.43 & 55499.43 & 18.92 & $16.673 \pm 0.012$ & i \\
2010-10-31.31 & 55500.31 & 19.80 & $16.687 \pm 0.011$ & i \\
2010-10-31.43 & 55500.43 & 19.92 & $16.674 \pm 0.012$ & i \\
2010-11-01.28 & 55501.28 & 20.77 & $16.697 \pm 0.012$ & i \\
2010-11-01.43 & 55501.43 & 20.92 & $16.688 \pm 0.012$ & i \\
2010-11-02.28 & 55502.28 & 21.77 & $16.699 \pm 0.012$ & i \\
2010-11-02.42 & 55502.42 & 21.91 & $16.685 \pm 0.010$ & i \\
2010-11-03.28 & 55503.28 & 22.77 & $16.697 \pm 0.013$ & i \\
2010-11-03.41 & 55503.41 & 22.90 & $16.678 \pm 0.012$ & i \\
2010-11-07.28 & 55507.28 & 26.77 & $16.691 \pm 0.017$ & i \\
2010-11-07.40 & 55507.40 & 26.89 & $16.688 \pm 0.010$ & i \\
2010-11-08.27 & 55508.27 & 27.76 & $16.695 \pm 0.014$ & i \\
2010-11-08.39 & 55508.39 & 27.88 & $16.692 \pm 0.014$ & i \\
2010-11-10.27 & 55510.27 & 29.76 & $16.689 \pm 0.012$ & i \\
2010-11-10.40 & 55510.40 & 29.89 & $16.689 \pm 0.014$ & i \\
2010-11-11.27 & 55511.27 & 30.76 & $16.702 \pm 0.014$ & i \\
2010-11-11.40 & 55511.40 & 30.89 & $16.699 \pm 0.011$ & i \\
2010-11-12.25 & 55512.25 & 31.74 & $16.714 \pm 0.017$ & i \\
2010-11-14.27 & 55514.27 & 33.76 & $16.710 \pm 0.014$ & i \\
2010-11-16.25 & 55516.25 & 35.74 & $16.706 \pm 0.012$ & i \\
2010-11-16.38 & 55516.38 & 35.87 & $16.685 \pm 0.010$ & i \\
2010-11-18.25 & 55518.25 & 37.74 & $16.707 \pm 0.010$ & i \\
2010-11-18.38 & 55518.38 & 37.87 & $16.684 \pm 0.011$ & i \\
2010-11-21.22 & 55521.22 & 40.71 & $16.711 \pm 0.013$ & i \\
2010-11-23.23 & 55523.23 & 42.72 & $16.697 \pm 0.013$ & i \\
2010-11-23.36 & 55523.36 & 42.85 & $16.686 \pm 0.013$ & i \\
2010-11-30.21 & 55530.21 & 49.70 & $16.701 \pm 0.019$ & i \\
2010-11-30.35 & 55530.35 & 49.84 & $16.659 \pm 0.025$ & i \\
2010-12-03.22 & 55533.22 & 52.71 & $16.662 \pm 0.046$ & i \\
2010-12-04.22 & 55534.22 & 53.71 & $16.691 \pm 0.011$ & i \\
2010-12-04.35 & 55534.35 & 53.84 & $16.688 \pm 0.011$ & i \\
2010-12-05.22 & 55535.22 & 54.71 & $16.690 \pm 0.016$ & i \\
2010-12-05.35 & 55535.35 & 54.84 & $16.677 \pm 0.017$ & i \\
2010-12-06.23 & 55536.23 & 55.72 & $16.691 \pm 0.016$ & i \\
2010-12-06.35 & 55536.35 & 55.84 & $16.666 \pm 0.029$ & i \\
2010-12-25.22 & 55555.22 & 74.71 & $16.675 \pm 0.020$ & i \\
2010-12-25.33 & 55555.33 & 74.82 & $16.662 \pm 0.022$ & i \\
2010-12-27.21 & 55557.21 & 76.70 & $16.673 \pm 0.016$ & i \\
2010-12-27.32 & 55557.32 & 76.81 & $16.686 \pm 0.017$ & i \\
2011-07-25.61 & 55767.61 & 287.10 & $16.717 \pm 0.015$ & i \\
2011-07-26.62 & 55768.62 & 288.11 & $16.727 \pm 0.012$ & i \\
2011-07-27.60 & 55769.60 & 289.09 & $16.721 \pm 0.013$ & i \\
2011-07-28.59 & 55770.59 & 290.08 & $16.735 \pm 0.018$ & i \\
2011-07-31.53 & 55773.53 & 293.02 & $16.729 \pm 0.018$ & i \\
2011-08-01.55 & 55774.55 & 294.04 & $16.725 \pm 0.011$ & i \\
2011-08-02.56 & 55775.56 & 295.05 & $16.714 \pm 0.010$ & i \\
2011-08-03.61 & 55776.61 & 296.10 & $16.727 \pm 0.010$ & i \\
2011-08-04.48 & 55777.48 & 296.97 & $16.677 \pm 0.015$ & i \\
2011-08-04.61 & 55777.61 & 297.10 & $16.779 \pm 0.019$ & i \\
\hline
\end{tabular}
\end{center}
\end{table}
%

%
\begin{table}[ht]
\addtocounter{table}{-1}
\caption{continued.}
\begin{center}
\begin{tabular}{lcccc}\hline\hline \noalign{\smallskip}
Date & MJD & $\Delta t ^a$ &  Mag. GC $^b$ & Filter\\
 {[}UT] & [d] & [d] & [mag] & \\ \hline \noalign{\smallskip}
2011-08-06.48 & 55779.48 & 298.97 & $16.696 \pm 0.011$ & i \\
2011-08-06.62 & 55779.62 & 299.11 & $16.710 \pm 0.011$ & i \\
2011-08-09.48 & 55782.48 & 301.97 & $16.693 \pm 0.020$ & i \\
2011-08-09.62 & 55782.62 & 302.11 & $16.742 \pm 0.014$ & i \\
2011-08-10.46 & 55783.46 & 302.95 & $16.685 \pm 0.010$ & i \\
2011-08-10.62 & 55783.62 & 303.11 & $16.724 \pm 0.009$ & i \\
2011-08-11.48 & 55784.48 & 303.97 & $16.713 \pm 0.012$ & i \\
2011-08-11.58 & 55784.58 & 304.07 & $16.696 \pm 0.012$ & i \\
2011-08-12.49 & 55785.49 & 304.98 & $16.702 \pm 0.020$ & i \\
\hline
\end{tabular}
\end{center}
\end{table}

\clearpage

\end{document}